\documentclass[10pt,journal,a4paper]{IEEEtran}
\usepackage{cite}      
\usepackage[caption=false, font=footnotesize]{subfig}

\usepackage[pdftex]{graphicx}

\usepackage[latin1]{inputenc}
\usepackage{url}       
\usepackage{amsfonts}
\usepackage{amsmath}
\usepackage{times}
\usepackage{algpseudocode}
\usepackage{algorithm}
\usepackage{enumerate}
\usepackage{siunitx}
\usepackage{multirow}
\usepackage{tikz}
\usepackage{color}

\begin{document}
\title{\hspace*{-0.1mm}Resource Allocation Frameworks for Network-coded Layered Multimedia Multicast Services}

\author{Andrea Tassi, Ioannis Chatzigeorgiou and Dejan Vukobratovi\'c
\thanks{This work is part of the R2D2 project, which is supported by EPSRC under Grant EP/L006251/1. Collaboration of the authors was facilitated by COST Action IC1104 on Random Network Coding and Designs over $\mathrm{GF}(q)$.

A. Tassi and I. Chatzigeorgiou are with the School of Computing and Communications, Lancaster University, Lancaster, UK (email: \{a.tassi, i.chatzigeorgiou\}@lancaster.ac.uk). 

D. Vukobratovi\'c is with the Department of Power, Electronics and Communication Engineering, University of Novi Sad, Serbia (email: \mbox{dejanv@uns.ac.rs}).}
}

\maketitle

\begin{abstract} 
The explosive growth of content-on-the-move, such as video streaming to mobile devices, has propelled research on multimedia broadcast and multicast schemes. Multi-rate transmission strategies have been proposed as a means of delivering layered services to users experiencing different downlink channel conditions. In this paper, we consider Point-to-Multipoint layered service delivery across a generic cellular system and improve it by applying different random linear network coding approaches. We derive packet error probability expressions and use them as performance metrics in the formulation of resource allocation frameworks. The aim of these frameworks is both the optimization of the transmission scheme and the minimization of the number of broadcast packets on each downlink channel, while offering service guarantees to a predetermined fraction of users. As a case of study, our proposed frameworks are then adapted to the LTE-A standard and the eMBMS technology. We focus on the delivery of a video service based on the H.264/SVC standard and demonstrate the advantages of layered network coding over multi-rate transmission. Furthermore, we establish that the choice of both the network coding technique and resource allocation method play a critical role on the network footprint, and the quality of each received video layer.
\end{abstract}


\section{Introduction}\label{sec:intro}
Multimedia multicast services will soon become a challenging issue to network service providers due to the increasing volume of multimedia traffic. Video content delivery represented $53$\% of the global mobile Internet traffic in 2013 and is expected to rise to $67$\% by 2018~\cite{CVI}. Considering the recent developments in fourth generation (4G) communication networks, a notable fraction of multimedia services is anticipated to be delivered over cellular networks. As the number of users increases, Point-to-Multipoint (PtM) transmission of a multimedia service is the natural choice over multiple and independent Point-to-Point (PtP) sessions. For this reason, 4G cellular networks have native support for broadcasting and multicasting services~\cite{6353684}. Recent work proposes to exploit this attractive inherent feature of 4G networks for broadcasting next generation Digital Television (DTV) services~\cite{6823652}. Furthermore, service multicasting over 4G infrastructures could also be used to deliver extra content in event locations, such as instant replays in sport venues~\cite{6730658}.

When a multicast service is transmitted by means of a single PtM data stream, the transmitting node sends the same data stream to all users. Given that users most likely experience heterogeneous propagation conditions, the transmission rate cannot be optimized for each user. Multirate Transmission (MrT) strategies overcome this issue by allowing users to recover different versions of the same PtM service~\cite{911162}. This paper focuses on MrT strategies that are suitable for \emph{layered services}~\cite{6397574}. A layered service consists of a \emph{base layer} and multiple \emph{enhancement layers}. The base layer allows each user to achieve a basic service quality, which is improved by using information conveyed by the enhancement layers. The $\ell$-th enhancement layer can be used to improve the service quality of a user only if both the base and the first $\ell-1$ enhancement layers have been successfully received by that user. In that context, a MrT strategy adapts the rate of each service layer by taking into account the heterogeneous propagation conditions between the transmitting node and the users.

The main goal of the considered family of MrT strategies is the maximization of the service level experienced by each user~\cite{6148193}. Most proposals divide users into multiple subgroups based on the user propagation conditions; each subgroup will eventually recover a different number of enhancement layers, in addition to the base layer. For example,~\cite{4917957,5452675} propose MrT strategies which achieve the aforementioned goal by maximizing the sum of service layers recovered by each user. However, little attention has been paid to the definition of MrT strategies which can ensure that specific subsets of layers will be recovered by predetermined fractions of users.

Our work relies on the MrT principle but proposes resource allocation frameworks that differ from those in the literature in terms of the achieved goal. In particular, we have restated the MrT resource allocation problem from the point of view of the network service provider; we have chosen as the optimization goal the minimization of the total amount of required radio resources to deliver a PtM layered service. Furthermore, owing to the idea of a service-level agreement between the service provider and the cell users, the constraint sets of the proposed optimization frameworks ensure that at least a predetermined fraction of users shall recover a given number of service layers with a target probability. A key point in the proposed MrT frameworks is that reliability of PtM communications is improved by means of the Random Linear Network Coding (RLNC) principle~\cite{Medard}. In particular, the resource allocation goal is fulfilled by jointly optimizing both the transmission parameters and the employed RLNC scheme.

\subsection{Related Works and Paper Contributions}
In our system model, each service layer forming a PtM service is delivered over multiple orthogonal broadcast erasure subchannels. Even though Automatic Repeat-reQuest (ARQ)~\cite{4441773} and Hybrid ARQ error control protocols~\cite{KiJiKSSc10} are suitable for broadcast erasure channels, the required amount of user feedback becomes intractable as the number of users grows. In order to mitigate this issue, reliability of multicast communications can be improved by means of Application Level-Forward Error Correction (AL-FEC) techniques, for example schemes based on Luby transform or low-density parity-check codes~\cite{6353684}. Unfortunately, as noted by E. Magli~\textit{et al.}~\cite{6416071}, this family of codes is designed to be applied over long source messages and, consequently, it introduces delay which is often undesirable in multimedia communications. In order to tackle this problem, several works propose the adoption of RLNC over \mbox{one-hop} broadcast networks~\cite{Ghaderi,MedardCap9,ITA}. A key point of RLNC implementations is that short source messages are preferred in order to reduce the decoding complexity and subsequently reduce the communication delay. Furthermore, various RLNC schemes for smartphones and low-end devices have been recently proposed, demonstrating that RLNC strategies are also affordable from the computational complexity point of view~\cite{6774596,6691231}. For these reasons, our work adopts the RLNC principle to address the reliability issues of PtM layered service transmissions.

Since each layer of a service has a different importance level, Unequal Error Protection (UEP) can be used to link the level of importance that a service layer has to the required level of protection. The UEP concept has been frequently applied to FEC schemes, see for example Reed-Solomon or low-density parity-check codes~\cite{6714525,4560155}, but was later adapted for RLNC codes~\cite{6168183}. This paper deals with two different UEP RLNC schemes~\cite{6168183}: the \emph{Non-Overlapping Window} (NOW-RLNC) and the \emph{Expanding Window} RLNC (EW-RLNC). Coded packets associated with a service layer $\ell$ are generated from source packets of \emph{layer $\ell$ only} in the case of NOW-RLNC or from source packets of \emph{the first $\ell$ layers} in the case of EW-RLNC.

Various resource allocation strategies have been proposed to support the transmission of network-coded multimedia flows over unreliable networks~\cite{R2,R5,R3,R4}. In particular,~\cite{R2} considers a system model where several \emph{single-layer} multimedia flows are broadcast to users forming a wireless mesh network. Each user linearly combines those incoming flows that can be decoded by other neighbouring users. Similarly to~\cite{R2}, the system model presented in~\cite{R5} is also concerned with a mesh network disseminating multimedia flows. However ~\cite{R5} considers layered multimedia streams whose reliability is improved by optimizing a distributed UEP RLNC implementation. In that case, each node realizes the UEP principle such that flows with high importance are more likely to be involved in linear combination operations. Differently to~\cite{R2,R5}, a two-hop content delivery network is studied in~\cite{R3}. The source node applies network coding to combine packets that form a layered multimedia service. The coded packets are then stored into several intermediate nodes. Subsequently, a single destination node retrieves the coded packets by connecting to the intermediate nodes via independent PtP sessions. According to the proposed UEP RLNC strategy in~\cite{R3}, which is valid for binary finite fields only, network-coded packets related to low-importance layers may depend on high-importance layers. Contrary to~\cite{R2,R5,R3},~\cite{R4} refers to a cellular network model, where the source node is in charge of generating and transmitting network coded packets to a \emph{single user}. The user acknowledges successfully received packets to the source node. If the acknowledged message is not received, either the same or a new coded packet is transmitted. The core idea of~\cite{R4} is that of optimizing the encoding process to minimize the total number of transmissions in a single PtP multimedia session.

In contrast to~\cite{R2,R5,R3}, our work refers to a typical cellular network topology, where the network coding operations are performed by the source node. Furthermore, this paper aims to jointly optimize the network coding process and the transmission parameters. In this way, we can view the RLNC implementation as a component which is fully integrated into the link adaptation framework of our communication system. Our proposal differs from~\cite{R4} both in terms of the considered RLNC strategies and the nature of the delivered data streams. More specifically,~\cite{R4} does not consider layered video services and, hence, does not investigate UEP RLNC strategies. Furthermore, the fact that the proposed scheme in~\cite{R4} has not been integrated into a more generic link adaptation framework hinders its extensibility to the case of PtM services.

Our analysis refers to a generic cellular network model, in a purely standard-independent fashion. However, in order to demonstrate the practical value of the proposed resource allocation frameworks, we present a case study, which refers to the 3GPP Long Term Evolution-Advanced (\mbox{LTE-A}) standard. The proposed implementation shows how our resource allocation frameworks can be adopted for the delivery of multimedia multicast services over the existing and, by following the same implementation guidelines, how can be also extended to next-generation networks. 

\mbox{LTE-A} integrates the evolved Multimedia Broadcast Multicast Service (eMBMS) framework, which enables it to handle multicast and broadcast services~\cite{sesia2011lte}. In the proposed implementation, we refer to multimedia multicast services that make use of the widely used H.264 video encoding standard and its scalable extension, known as Scalable Video Coding (H.264/SVC), which is gaining popularity~\cite{h264}. In line with our considered layered message structure, the H.264/SVC encoder transforms a raw video stream into a layered service, such that \emph{enhancement layers} improve the resolution of a \emph{base video layer} of a stream~\cite{6025326}. In order to make the considered network-coded service delivery suitable for multicasting over an \mbox{LTE-A} network, we have adopted the proposal of integrating a RLNC encoder into the \mbox{LTE-A} protocol stack, as described in~\cite{6353397}. In its original version, the proposed integration refers to a system model according to which a \emph{PtP data stream} is transmitted by a base station to a \emph{single user}, either directly or via a relay node. The system design proposed in~\cite{6353397} was later enhanced in~\cite{TVTTassi} in order to broadcast H.264/SVC video streams as eMBMS flows. Concerning the optimization frameworks that will be presented, this work builds on and extends the idea presented in~\cite{TVTTassi}. In particular,~\cite{TVTTassi} provides a resource allocation model minimizing the total number of transmission attempts needed to broadcast a H.264/SVC video stream. Even though we aim at fulfilling the same objective, this paper significantly differs to~\cite{TVTTassi} in terms of the considered radio resource model. We refer to a generic system model where coded packets are transmitted over a set of orthogonal subchannels. Unlike~\cite{TVTTassi}, we develop resource allocation frameworks which \emph{allow} coded packets associated with different video layers to be mixed within the same subchannel to enhance user performance, both in the case of NOW- and EW-RLNC. For any of the proposed resource allocation models, we provide efficient heuristic strategies capable of finding a good quality resource allocation solution in a finite number of steps.

With regards to the coding schemes that we will refer to, unlike~\cite{6353397} and~\cite{TVTTassi}, this work focuses on NOW- and \mbox{EW-RLNC} schemes suitable for layered service transmissions. In addition, the authors of~\cite{6353397,TVTTassi} did not optimize the bit length of source packets used to represent the transmitted layered service; the source packet bit length is given a priori. This paper proposes a model for optimizing the source packet bit length to fit the transmission constraints of the communication standard in use. Since the bit length of source packets is constrained to be smaller than or equal to a maximum target value, the number of source packets representing a layered service can be upper-bounded. Hence, this work can represent the same layered service with a smaller number of source packets, compared to what proposed in~\cite{TVTTassi}. We remark that the number of source packets has a significant impact on the computation complexity of the RLNC decoding phase~\cite{Medard}.

The remaining part of the paper is organized as follows. In Section~\ref{sec:system}, we present the considered standard-independent system model and derive the necessary theoretical foundations needed to assess the performance of NOW- and EW-RLNC. Section~\ref{subsec:raModels} builds upon the aforementioned system model the proposed resource allocation models suitable for optimizing layered multicast communications. Section~\ref{sec:eMBMS} shows, as a case study, how the proposed modelling and resource allocation frameworks can be implemented in a practical communication system, such as \mbox{LTE-A}. Analytical results are discussed in Section~\ref{sec:results}, whereas Section~\ref{sec:conclusions} summarizes the main findings of the paper.

\section{System Parameters and Performance Analysis}\label{sec:system}
\begin{table}[tbd]
\centering
\caption{Commonly used notation.}
\label{tab.not}
{\scriptsize
\begin{tabular}{|c|p{6.53cm}|}
\hline \multirow{ 2}{*}{$\Hat{B}_c$}  &	Maximum number of coded packets that can be transmitted over\\
                    &	subchannel $c$\\
\hline $m_c$  & Modulation and Coding Scheme (MCS) adopted by subchannel $c$\\
\hline \multirow{ 2}{*}{$p_u(m)$}  & Packet Error Rate (PER) of user $u$ when MCS with index $m$ is\\
                 & adopted\\
\hline $\Hat{p}$ &	The reception of a coded packet is acceptable if $p_{u}(m) \leq \Hat{p}$\\
\hline $p_u$  &	Defined in~\eqref{eq.p} and approximated by~\eqref{eq.pLTE} \\
\hline $M(u)$ &	The greatest value of $m$ for which $p_{u}(m) \leq \Hat{p}$ \\
\hline $H$  & Source/coded packet bit length	\\
\hline \multirow{ 2}{*}{$\mathbf{x}$}  & Layered source message that consists of $K$ equal-length source	\\
                     & packets	\\
\hline $\mathbf{w}_\ell$  & Set of $k_\ell$ source packets composing the $\ell$-th layer	\\
\hline $\mathbf{W}_\ell$  &	Set of $K_\ell$ source packets belonging to the first $\ell$ service layers\\
\hline \multirow{ 2}{*}{$n^{(\ell,c)}$}  & Number of coded packets related to layer $\ell$ layer and transmitted\\
                       & over subchannel $c$	\\
\hline $\mathbf{n}_u$  & Vector $\{n_{1,u}, \ldots, n_{L,u}\}$, where $n_{\ell,u}$ is defined by~\eqref{eq.nlu}\\
\hline $\mathbf{N}_u$  & Vector $\{N_{1,u}, \ldots, N_{L,u}\}$, where $N_{\ell,u}$ is defined by~\eqref{eq.N_lu}	\\
\hline \multirow{ 2}{*}{$\mathrm{P}^{\mathrm{NOW}}_{1:\ell}(\mathbf{n}_u)$}  & Probability that user $u$ will recover the first $\ell$ service layers, in \\
& the case of the NOW-RLNC\\
\hline \multirow{ 2}{*}{$\mathrm{P}^{\mathrm{EW}}_{1:\ell}(\mathbf{N}_u)$}  &	Probability of user $u$ recovering the $\ell$-th window, in the case of\\ 
& the EW-RLNC\\
\hline \multirow{ 2}{*}{$\mathrm{P}^{\mathrm{MrT}}_{1:\ell}$}  & Probability of user $u$ recovering the $\ell$-th window, when the MrT\\
& strategy is in use\\
\hline
\end{tabular}
}
\end{table}

We consider an one-hop wireless communication system composed of one source node and $U$ users. Each transmitted data stream is delivered to users through $C$ orthogonal broadcast erasure subchannels. In our system model we have that all the data streams are conveyed to the users according to the RLNC principle. As a consequence, each subchannel delivers streams of network-coded packets (henceforth referred to as \textit{coded packets} for brevity) that may be associated with one or more data streams. Furthermore, we impose that the maximum length of a stream, in terms of the number of coded packets that can be transmitted over the $c$-th subchannel during a given time interval, for $c = 1, \ldots, C$, is fixed and equal to $\Hat{B}_c$. In particular, we assume that indexes $c=1, 2, \ldots, C$ are assigned to subchannels so that the relation $\Hat{B}_1 \leq \Hat{B}_2 \leq \ldots \leq \Hat{B}_C$ holds. For clarity, Table~\ref{tab.not} summarizes the symbols commonly used in the paper.

Each element of a coded packet stream is delivered by means of a specific Modulation and Coding Scheme (MCS), which is identified by nonnegative integer $m$. We denote by $p_{u}(m)$ the Packet Error Rate (PER) that a user $u$ experiences  when $m$ is the index of the adopted MCS. If $m^\prime$ and $m^{\prime\prime}$ are indexes identifying two different MCSs and $m^\prime \leq m^{\prime\prime}$, then the MCS described by $m^{\prime\prime}$ either uses a higher modulation order or reduced error-correcting capability than the MCS represented by $m^{\prime}$. Naturally, for the same channel conditions, it follows that $p_{u}(m^\prime) \leq p_{u}(m^{\prime\prime})$ also holds. In general, we regard reception of a coded packet as being \emph{acceptable} if $p_{u}(m)$ is equal to or smaller than a predetermined threshold $\Hat{p}$. To this end, if user $u$ can choose from a range of MCSs, we define $M(u)$ as the greatest value of $m$ for which $p_{u}(m) \leq \Hat{p}$, \mbox{that is}
\begin{equation}
M({u})\doteq\{\;m\;\;|\;\; p_{u}(m) \leq \Hat{p} \;\; \wedge \;\; p_{u}(m+1) > \Hat{p}\;\}.\label{eq.CQI}
\end{equation}

In the system model presented in this paper, we also impose that coded packets transmitted through the $c$-th subchannel shall use the same MCS, characterized by index $m_c$. As will become evident in the rest of the paper, the determination of the optimal MCS for each subchannel, $m_1, \ldots, m_C$, is part of the proposed resource allocation strategies.

Let $\mathbf{x} = \{x_1, \ldots, x_K\}$ be a layered source message that consists of $K$ equal-length source packets, classified into $L$ service layers. For simplicity and without loss of generality, we assume that packets in the source message are arranged in order of decreasing importance. In other words, the first service layer appears at the beginning of the source message and is followed by progressively less important layers, until the last and least important $L$-th service layer. If the $\ell$-th layer consists of $k_\ell$ data packets, we observe that $K = \sum_{\ell = 1}^L k_\ell$. Throughout this paper, we define the Quality-of-Service (QoS) level experienced by a user as the number of \emph{consecutive source message layers} that can be recovered, starting from the first layer.

In the remainder of this section, we present the layered RLNC strategies under consideration. In addition we provide accurate models to evaluate the probability that a source message transmitted by means of NOW-RLNC and EW-RLNC is correctly received by a user. Theoretical results discussed in the rest of this section are general and apply to any cellular system model, where: (i) data flows can be delivered by using different MCSs, and (ii) each source message layer is broadcast through independent communication subchannels.

\subsection{Performance of Non-Overlapping Window RLNC}\label{subsec:NO_Rnc}
We first consider the case where the source node uses the RLNC principle on each individual layer of the source message. Let us define $K_\ell$ as $K_\ell =\sum_{i=1}^{\ell}k_i$. The source node will linearly combine the $k_\ell$ data packets composing the $\ell$-th layer \mbox{$\mathbf{w}_\ell=\{x_i\}_{i=K_{\ell-1} + 1}^{K_\ell}$} and will generate a stream of $n_\ell\geq k_\ell$ coded packets $\mathbf{y}=\{y_j\}_{j=1}^{n_\ell}$, where $y_j = \sum_{i=K_{\ell-1}+1}^{K_{\ell}} g_{j,i} \cdot x_i$. Coefficient $g_{j,i}$ is uniformly selected at random over a finite field $\mathbb{F}_q$ of size $q$. We refer to this encoding strategy as NOW-RLNC throughout this paper.

A stream of coded packets associated with a service layer can be broadcast to the $U$ users over a single subchannel or multiple subchannels. Let $n^{(\ell,c)}$ signify the number of coded packets that are related to the $\ell$-th layer and are transmitted over the $c$-th subchannel. We expect that some or all of these $n^{(\ell,c)}$ coded packets will be received by user $u$, if the predetermined PER requirement is met, i.e. $M(u) \geq m_c$. Otherwise, user $u$ will not recover any of the $n^{(\ell,c)}$ coded packets. We can express the maximum number of coded packets associated with the $\ell$-th layer that user $u$ can collect from the $C$ subchannels as
\begin{equation}
n_{\ell,u} = \mathop{\sum_{c = 1}^C} n^{(\ell,c)}\;I\left(M(u) \geq m_c\right) \label{eq.nlu}
\end{equation}
where $I(\cdot)$ is an indicator function where $I(\cdot)=1$ if its input argument is true, otherwise $I(\cdot)=0$.

To simplify our analysis, we introduce $p_u$ as the maximum PER value experienced by user $u$ across all subchannels that offer acceptable reception and convey at least one coded packet (namely, $\sum_{\ell = 1}^{L} n^{(\ell,c)} > 0$), that is
\begin{equation}
p_u \doteq \!\!\max_{c = 1, \ldots, C}\!\left(p_{u}(m_c) \,\,\Big|\,\, M(u) \geq m_c \wedge \sum_{\ell = 1}^{L} n^{(\ell,c)} > 0\right).\! \label{eq.p}
\end{equation} 
Based on~\cite{6550868}, we can infer that if $n_{\ell,u}$ coded packets are transmitted over those subchannels such that $M(u) \geq m_c$, user $u$ will recover the $\ell$-th layer with probability 
\begin{equation}
\mathrm{P}^{\mathrm{NOW}}_{\ell}(n_{\ell,u}) = \displaystyle\sum_{r = k_\ell}^{n_{\ell,u}} \mathrm{P}_{\mathrm{R}}(n_{\ell,u},r) \,\, \mathrm{P}_{\mathrm{D},\ell}(r) \label{eq.FC}
\end{equation}
where
\begin{equation}
\mathrm{P}_{\mathrm{R}}(n_{\ell,u},r) = \binom{n_{\ell,u}}{r} \, p_u^{n_{\ell,u}-r} \, (1-p_u)^{r}
\end{equation}
represents the probability that $r$ out $n_{\ell,u}$ coded packets are successfully  received by user $u$, when the PER is given by~\eqref{eq.p}. In addition, the term 
\begin{equation}
\mathrm{P}_{\mathrm{D},\ell}(r) = \prod_{i=0}^{k_\ell - 1} \Big[1 - \frac{1}{q^{r-i}}\Big] \label{eq.funcH}
\end{equation}
is the probability that $k_\ell$ out of $r$ received coded packets are linearly independent, i.e., $\mathrm{P}_{\mathrm{D},\ell}(r)$ is the probability that the source packets forming $\mathbf{w}_\ell$ can be recovered~\cite{5634159}. The joint probability that user $u$ will recover the first $\ell$ service layers, i.e. $1, 2, \ldots, \ell$, is simply the product of the $\ell$ individual probabilities, which can be written as
\begin{equation}
\mathrm{P}^{\mathrm{NOW}}_{1:\ell}(\mathbf{n}_u) = \prod_{i = 1}^{\ell} \mathrm{P}^{\mathrm{NOW}}_{i}(n_{i,u}) \label{eq.noFC}
\end{equation}
where $\mathbf{n}_u=\{n_{1,u}, \ldots, n_{L,u}\}$.

\begin{figure}[tbd]
\centering
\includegraphics[width=0.7\columnwidth]{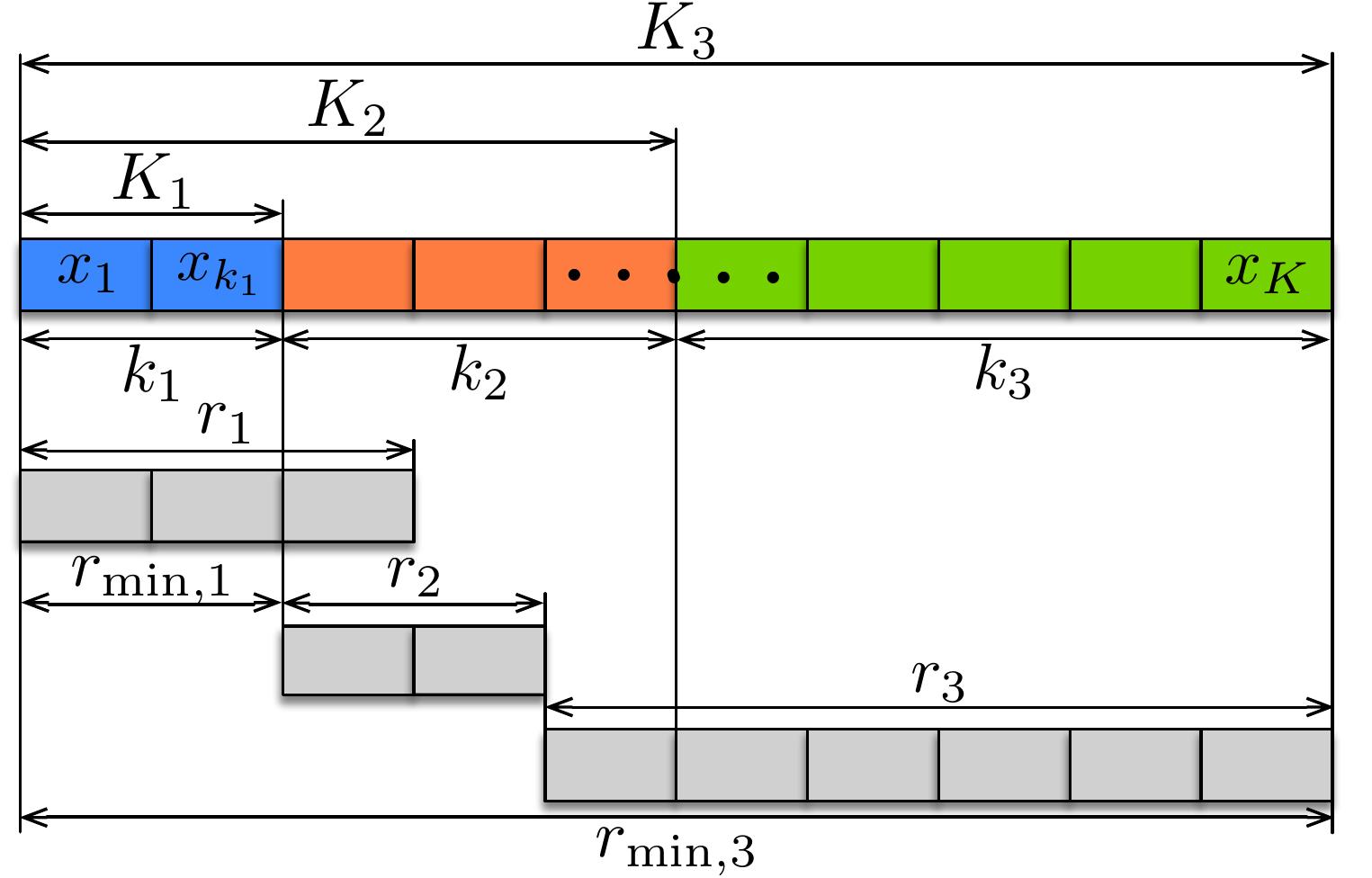}
\caption{Expanding Window source message model, a possible combination of coded packets that have been received and graphic interpretation of $r_{\mathrm{min},1}$ and $r_{\mathrm{min},3}$.}
\label{fig.fig_ew_model}
\end{figure}

\subsection{Performance of Expanding Window RLNC}\label{subsec:EW_Rnc}
We will now shift our focus onto a different RLNC approach known as the expanding window principle, which was presented in~\cite{6168183}. To this end, we consider the same layered source message $\mathbf{x}$ as before, and define the $\ell$-th window $\mathbf{W}_\ell$ as the set of source packets belonging to the first $\ell$ service layers. As depicted in Fig.~\ref{fig.fig_ew_model}, a window spanning over the first $\ell$ layers can be expressed as $\mathbf{W}_\ell = \{\mathbf{w}_i\}_{i=1}^{\ell}$ or, equivalently, $\mathbf{W}_\ell\!=\!\{x_j\}_{j=1}^{K_\ell}$. In the case of EW-RLNC, the source node (i) linearly combines data packets belonging to the same window, (ii) repeats this process for all windows, and (iii) broadcasts each stream of coded packets associated with each window over one or more subchannels.

In a similar fashion to the NOW case, we define $N^{(\ell,c)}$ as the number of coded packets that are associated with the \mbox{$\ell$-th} window and are transmitted over the $c$-th subchannel. The maximum number of coded packets related to the \mbox{$\ell$-th} window that user $u$ can collect through the $C$ subchannels is
\begin{equation}
N_{\ell,u} = \sum_{c = 1}^{C} N^{(\ell,c)}\;I\left(M(u) \geq m_c\right)\text{.} \label{eq.N_lu}
\end{equation}

Using~\eqref{eq.N_lu}, we can obtain vector $\mathbf{N}_u=\{N_{1,u},\ldots,N_{L,u}\}$, which describes the maximum number of transmitted coded packets, related to each window, that can be collected by user $u$. The objective of this section is to derive a \mbox{closed-form} expression for $\mathrm{P}^{\mathrm{EW}}_{1:\ell}(\mathbf{N}_u)$, which denotes the probability of user $u$ recovering the $\ell$-th window and thus retrieving the first $\ell$ layers. To do that, we shall first consider vector $\mathbf{r}=\{r_{1},\ldots,r_L\}$, which describes the number of successfully received coded packets associated with each window, and study the requirements for which $\mathrm{P}^{\mathrm{EW}}_{1:\ell}(\mathbf{N}_u)>0$.

For a given set of received coded packets $\mathbf{r}$, we define the minimum number of coded packets associated with the $\ell$-th expanding window, denoted as $r_{\min,\ell}$, which shall be successfully received such that the probability of recovering $\mathbf{W}_\ell$, by considering just coded packets associated with the first $\ell$ expanding windows, is non-zero. Clearly, for $\ell = 1$, we have that $r_{\min,1}=K_1$. Indeed, as per the properties of random network coding, the first window ($\ell=1$) is likely to be decoded ($\mathrm{P}^{\mathrm{EW}}_{1:\ell}(\mathbf{N}_u)>0$) only if: (i) the number of received coded packets pertaining to the first window is at least equal to the number of source packets comprising that window $(r_1\geq K_1)$, or (ii) 
the probability of recovering a larger window is greater than zero.

Consider Fig.~\ref{fig.fig_ew_model}, which provides a graphical interpretation of $r_{\min,1}$ and $r_{\min,3}$. In the reported example, given that \mbox{$r_{\min,1} = K_1$}, we note that $r_{\min,1} + r_2$ is less than $K_2$. Hence, the set of source packets $\mathbf{W}_2$ cannot be recovered because the number of linearly independent coded packets associated with the first two windows cannot be equal to $K_2$. However, in this case, the value of $r_3$ is such that $r_{\min,1} + r_2 + r_3$ is equal to $K_3$. This means that the probability of having $K_3$ linearly independent coded packets and recovering $\mathbf{W}_3$ is greater than zero. We also note that, in the considered example, the value of $r_3$ is the smallest one such that $\mathrm{P}^{\mathrm{EW}}_{1:3}(\mathbf{N}_u)>0$ holds. Hence, $r_{\min,3} = r_{\min,1} + r_2 + r_3$. In general, the remaining values of $r_{\min,\ell}$, for $\ell = 2, \ldots, L$, can be computed using the following recursion:
\begin{equation}
r_{\min,\ell} = K_\ell - K_{\ell-1} + \max\left(r_{\min,\ell-1}\:-\:r_{\ell-1},\:\:0\right)
\end{equation}
which asserts that the probability of decoding the first $\ell$ layers is non-zero if the number of received coded packets related to the $\ell$-th window is at least equal to the size difference between windows $\ell$ and $\ell-1$, complemented by a possible packet deficit carried over from window $\ell\!-\!1$.

Having derived an expression for $r_{\min,\ell}$, for $N_{\ell,u} > 0$, the probability of user $u$ recovering the first $\ell$ layers, $\mathrm{P}^{\mathrm{EW}}_{1:\ell}(\mathbf{N}_u)$, can be written as the probability $\mathrm{P}^{\mathrm{EW}}_{1:\ell}(\mathbf{N}_u,\mathbf{r})$ of successfully receiving $r = \{r_t\}_{t=1}^{\ell}$ coded packets \textit{and} recovering the $\ell$-th window, summed over all valid values of $\mathbf{r}$. In other words, we can write
\begin{equation}
\label{eq.Pd_ew_complete}
\mathrm{P}^{\mathrm{EW}}_{1:\ell}(\mathbf{N}_u)= 
\sum_{r_1 = 0}^{N_{1,u}} \!\!\cdots\!\!\!\!\sum_{r_{\ell\!-\!1} = 0}^{N_{\ell\!-\!1,u}}\:\sum_{r_\ell = r_{\min,\ell}}^{N_{\ell,u}}\!\!\!\mathrm{P}^{\mathrm{EW}}_{1:\ell}(\mathbf{N}_u,\mathbf{r}).
\end{equation}
Let
\begin{equation}
\label{eq.Pd_ew_part1}
\mathrm{P}_{\mathrm{R}}(\mathbf{N}_u,\mathbf{r}) = \prod_{i = 1}^{\ell} \binom{N_{i,u}}{r_i} p_u^{N_{i,u}-\,r_i} \left(1-p_u\right)^{r_i}
\end{equation}
be the probability of receiving $r_i$ out of $N_{i,u}$ coded packets, where the PER is given by~\eqref{eq.p}, for any $i = 1, \ldots, \ell$. Of course, in this case, the term $n^{(\ell,c)}$ in~\eqref{eq.p} is replaced with $N^{(\ell,c)}$. The relation
\begin{equation}
\label{eq.Pd_ew_part2}
\mathrm{P}^{\mathrm{EW}}_{1:\ell}(\mathbf{N}_u,\mathbf{r}) = \mathrm{P}_{\mathrm{R}}(\mathbf{N}_u,\mathbf{r})\:\mathrm{P}_{\mathrm{D},1:\ell}(\mathbf{r})
\end{equation}
considers all possible combinations of receiving $\mathbf{r}$ coded packets out of $\mathbf{N}_u$ packets, multiplied by the probability $\mathrm{P}_{\mathrm{D},1:\ell}(\mathbf{r})$ of successfully decoding the source message $\mathbf{W}_\ell$. Similarly to~\eqref{eq.funcH}, $\mathrm{P}_{\mathrm{D},1:\ell}(\mathbf{r})$ represents the probability of having $K_\ell$ linearly independent coded packets out of the $\sum_{i = 1}^{\ell} r_i$ received ones.

Owing to the lack of an accurate expression for $\mathrm{P}_{\mathrm{D},1:\ell}(\mathbf{r})$, we approximated it by using~\eqref{eq.funcH}. Let $\mathbf{r}^\prime = \{r^\prime_t\}_{t = 1}^{L}$ be a vector of $\ell$ elements, where $r^\prime_t = 0$ if $t \neq \ell$, otherwise $r^\prime_\ell = \sum_{j=1}^{\ell}r_j$. It is straightforward to note that the relation $\mathrm{P}_{\mathrm{D},1:\ell}(\mathbf{r}) \leq \mathrm{P}_{\mathrm{D},1:\ell}(\mathbf{r}^\prime)$ holds. In addition, from~\eqref{eq.funcH}, we understand that $\mathrm{P}_{\mathrm{D},1:\ell}(\mathbf{r}^\prime)$ is equal to $\mathrm{P}_{\mathrm{D},\ell}\left(r^\prime_\ell\right)$. For these reasons, we decide to approximate $\mathrm{P}_{\mathrm{D},1:\ell}(\mathbf{r})$ as follows:
\begin{equation}
\mathrm{P}_{\mathrm{D},1:\ell}(\mathbf{r}) \simeq \mathrm{P}_{\mathrm{D},\ell}\!\!\left(\sum_{j=1}^{\ell}r_j\right) \!=\!\!\! \prod_{i=0}^{K_\ell - 1} \left[1 - \frac{1}{q^{\left(\sum_{j=1}^{\ell}r_j\right)-i}}\right]\text{.}\!\label{eq.gFunc}
\end{equation}

\begin{figure}[tbd]
\vspace{-2.8mm}\centering
\subfloat[$q = 2$]{\label{fig.val2}
\hspace{-1.3mm}\includegraphics[width=0.5\columnwidth]{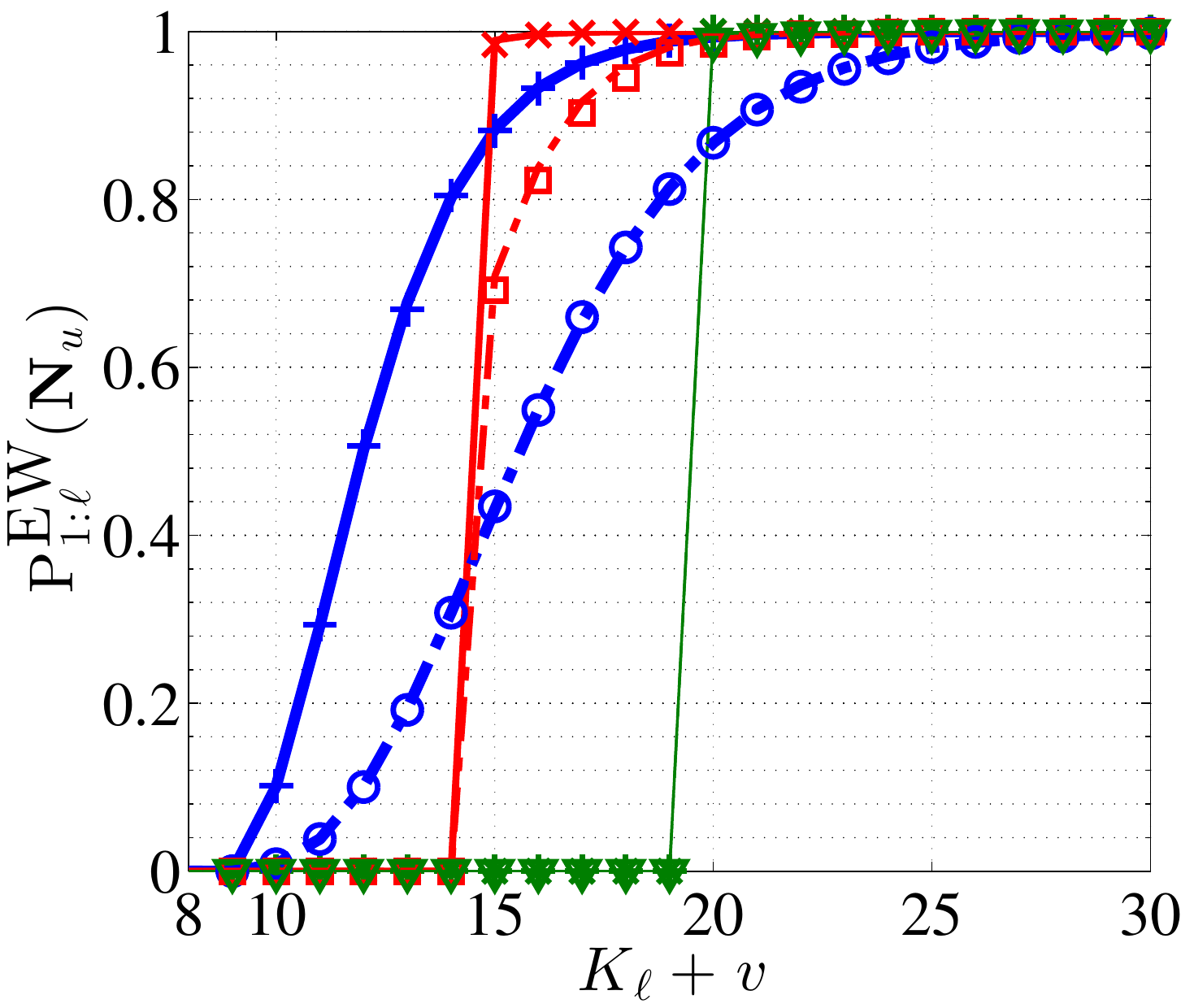}
}
\subfloat[$q = 2^8$]{\label{fig.val28}
\hspace{-2mm}\includegraphics[width=0.5\columnwidth]{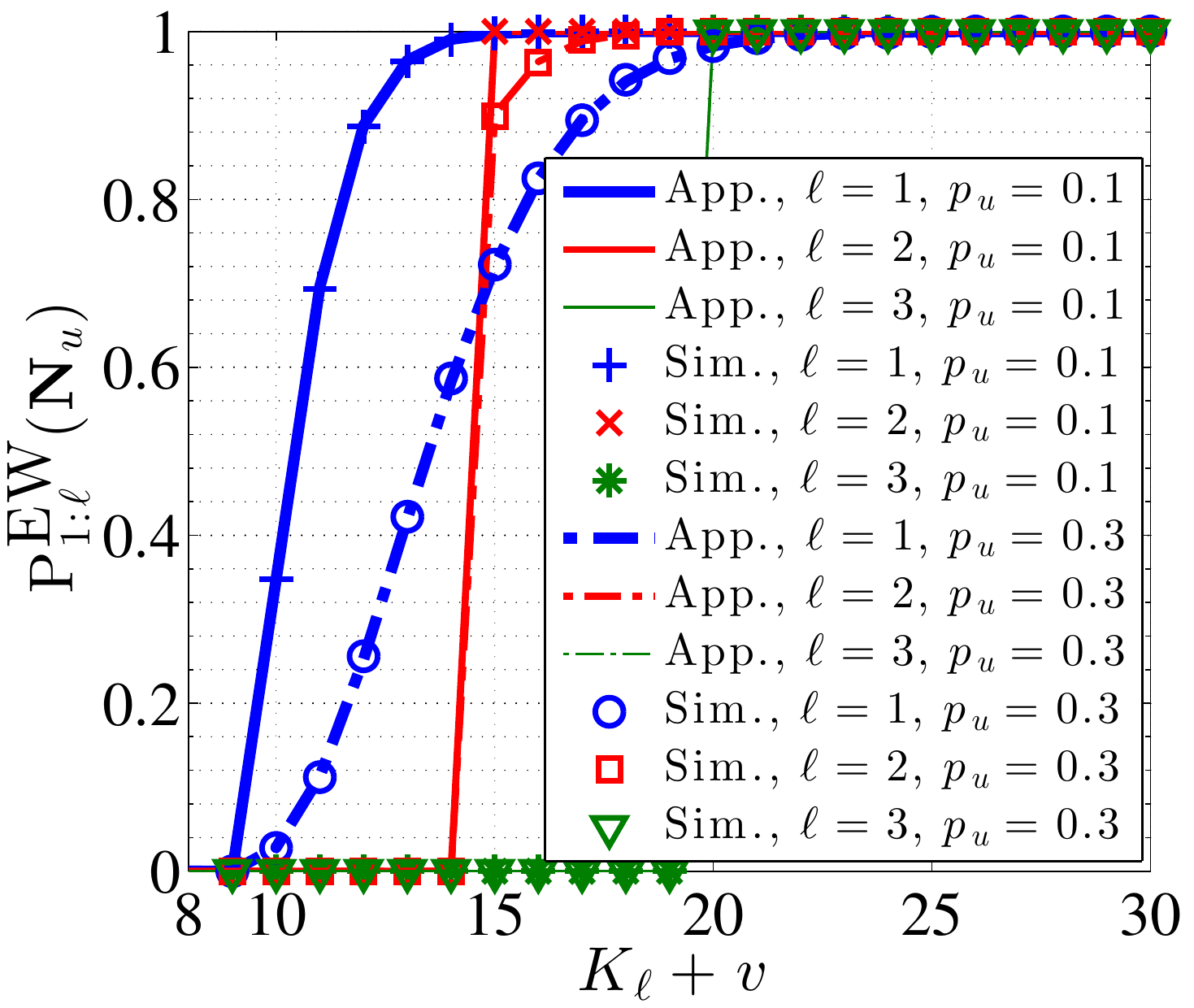}
}
\caption{Performance comparison between the approximated and simulated version of $\mathrm{P}^{\mathrm{EW}}_{1:\ell}(\mathbf{N}_u)$, refer to the legend of Fig.~\ref{fig.val28} for both figures.}
\label{fig.val}
\end{figure}

In order to inspect the quality of the considered approximation, we compared probabilities obtained by using~\eqref{eq.Pd_ew_complete} with those obtained by computer simulations, for different values of $p_u = 0.1$ or $0.3$ and finite field sizes $q = 2$ or $q = 2^8$. In particular, Fig.~\ref{fig.val} compares both the approximated and the simulated value of $\mathrm{P}^{\mathrm{EW}}_{1:\ell}(\mathbf{N}_u)$, where $N_{\ell,u} = K_\ell + v$, for $\ell = 1, \ldots, 3$ and $v \geq 0$. We consider $K_1 = 5$, $K_2 = 10$ and $K_3 = 15$. Note that the maximum performance gap between the approximated and the simulated results occurs for $p_u = 0.3$ and it is smaller than $0.017$ for $q=2$, and $0.004$ for $q=2^8$. The performance gap between approximated and simulated results becomes negligible for an increasing value of $q$.

\section{Proposed Multi-Channel Resource Allocation Models and Heuristic Strategies}\label{subsec:raModels}
In this section, we propose strategies that can be used to allocate coded packets over the set of communication subchannels. All the proposed optimization models jointly optimize the MCSs associated with each subchannel and the number of coded packet transmissions. The objective of the proposed models aim at minimizing the total number of coded packet transmissions needed to deliver service layers. This minimization is constrained by the fact that (at least) a predetermined fraction of users shall be able to recover a given set of service layers with (at least) a target probability. For each proposed optimization model, efficient heuristic strategies are provided.

\begin{figure}[tbd]
\centering
\subfloat[Separated allocation pattern]{\label{fig.SP}
		\hspace{-2.4mm}\includegraphics[width=0.5\columnwidth]{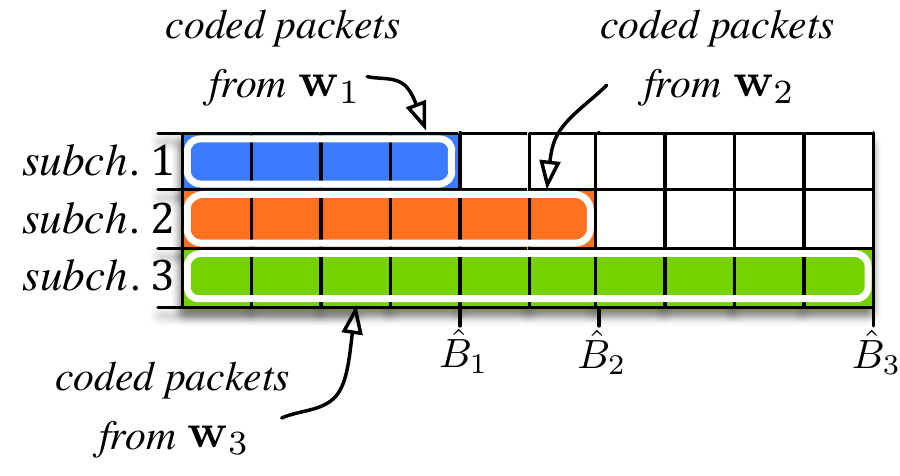}
}
\subfloat[Mixed allocation pattern]{\label{fig.MP}
		\hspace{-1.3mm}\includegraphics[width=0.5\columnwidth]{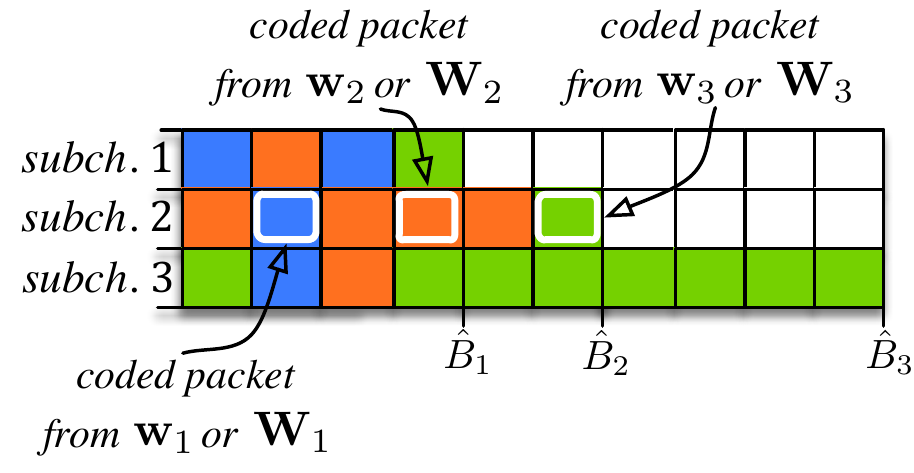}
}
\caption{Considered radio allocation patterns, for $C = 3$ subchannels (``subch.'') and $L = 3$ layers.}
\label{fig.patterns}
\end{figure}

Before going into the details of the proposed resource allocation models, we consider the following allocation patterns:
\begin{itemize}
\item Separated Allocation (SA) pattern (Fig.~\ref{fig.SP}), where a stream of coded packets associated with a service layer or expanding window shall be mapped on one subchannel only. This means that coded packets belonging to different layers or windows cannot be mixed within the same subchannel. 
\item Mixed Allocation (MA) pattern (Fig.~\ref{fig.MP}), where coded packets belonging to different service layers or windows can be delivered through the same subchannel.
\end{itemize}
In this section we refer to the generic system model described in Section~\ref{sec:system}. Hence, the resource allocation frameworks that will be presented are also generic and standard-independent.

\subsection{Non-Overlapping Window Resource Allocation Strategies}\label{NO}
Consider a system where the source node delivers the layered service by means of the NOW-RLNC principle. From~\eqref{eq.noFC}, we define the indication variable $\lambda_{u,\ell}$ as follows:
\begin{equation}
\lambda_{u,\ell} = I\left(\mathrm{P}^{\mathrm{NOW}}_{1:\ell}(\mathbf{n}_u) \geq \Hat{P}\right).\label{eq.delta}
\end{equation}
In other words, $\lambda_{u,\ell} = 1$, if $u$ can recover the first $\ell$ layers with a probability value that is equal to or greater than a target value $\Hat{P}$, otherwise $\lambda_{u,\ell} = 0$. Equivalently, we can say that if $\lambda_{u,\ell} = 1$, $u$ achieves the QoS level $\ell$ with at least a probability of $\Hat{P}$.

The resource allocation model that we propose for the case of NOW-RLNC employing SA (NOW-SA) can be formulated as follows:
\begin{align}
	\text{(NOW-SA)} &  \quad  \mathop{\min_{m_1, \ldots, m_C}}_{n^{(1,c)}, \ldots, n^{(L,c)}} \,\,  \sum_{\ell=1}^{L} \sum_{c=1}^{C} n^{(\ell,c)} \label{NO-SA.of}\\
    \text{subject to} &   \quad \sum_{u = 1}^U \lambda_{u,\ell} \geq U \, \Hat{t}_\ell \,\,\,\,\quad\quad\text{$\ell = 1, \ldots, L$}\label{NO-SA.c1}\\
                      &   \quad m_{c-1} < m_{c} \quad\quad\quad\quad\, \text{$c = 2, \ldots, L$}\label{NO-SA.c2}\\
                      &   \quad 0 \leq \sum_{\ell=1}^{L} n^{(\ell,c)} \leq \Hat{B}_c \quad  \text{$c = 1, \ldots, C$}\label{NO-SA.c3}\\
                      &   \quad n^{(\ell,c)} = 0   \quad\quad\quad\quad\,\,\,\,\,\text{ for $\ell \neq c$}\label{NO-SA.c4}
\end{align}
where the objective function~\eqref{NO-SA.of} represents the overall number of coded packet transmissions needed to deliver all the $L$ service layers. Furthermore, constraint~\eqref{NO-SA.c1} ensures that the fraction of users that can recover the first $\ell$ service layers is equal to or greater than a desired value $\Hat{t}_\ell$. In order to let the model exploit user heterogeneity, constraint~\eqref{NO-SA.c2} avoids the situation in which two subchannels are transmitted using the same MCS. Constraint~\eqref{NO-SA.c3} ensures that the number of coded packets delivered by any subchannel does not exceed $\Hat{B}_c$. Constraint~\eqref{NO-SA.c4} avoids that coded packets associated with different service layers are mixed within the same subchannel. Hence, in this case, $C$ has to be equal to or greater than $L$.

Considering the case of a MA pattern, the service delivery based on the NOW-RLNC approach can be optimized by means of a new optimization model which we shall refer to as NOW-MA. This new optimization model has the same definition of the NOW-SA but, in this case, we remove constraint~\eqref{NO-SA.c4}. In this way, coded packets associated with different service layers can be delivered by means of the same subchannel and $L$ can be different from $C$.

Unfortunately, both the NOW-SA and NOW-MA are hard integer optimization problems because of constraints~\eqref{NO-SA.c1} and~\eqref{NO-SA.c3} that introduce strong coupling relations among delivered service layers. To this end, we propose a couple of two-step heuristic strategies suitable for deriving, in a finite number of iterations, good quality solutions for both aforementioned problems. In particular, the idea underlying each heuristic approach is that of separating the optimization of MCS (associated with each subchannel) from the number of coded packets (related to each service layer) to be delivered.

\begin{algorithm}[tbd]
\floatname{algorithm}{Procedure}
\caption{Subchannel MCSs optimization.}
\label{Alg.P1}
{\scriptsize \begin{algorithmic}[1]
\State $c \gets C$
\State $v \gets m_{\mathrm{MAX}}$ and 
\While {$c \geq 1$} 
	\Repeat
		\State $m_c \gets v$
		\State $v \gets v - 1$
	\Until {$|\mathcal{U}^{(m_c)}| \geq U \cdot \Hat{t}_c$ \textbf{or} $v < m_{\mathrm{min}}$}
	\State $c \gets c - 1$
\EndWhile
\end{algorithmic}}
\end{algorithm}

Considering the SA pattern, the first step of the proposed heuristic strategy aims at optimizing variables $m_c$, for \mbox{$c = 1, \ldots, C$}. In this case, the value of $C$ has to be equal to $L$ because of the nature of the considered allocation pattern. Furthermore, without loss of generality, we assume that the coded packet stream associated with layer $\ell$ is delivered by means of the \emph{$\ell$-th subchannel\footnote{To this end, in the case of the SA pattern we reference both subchannels and service layers with the same index $\ell$.}}. Let $\mathcal{U}^{(m_c)}$ be a set of users such that $u \in\mathcal{U}^{(m_c)}$ if $M(u) \geq m_c$. The first step of the heuristic aims at selecting the value of $m_c$ such that the cardinality of $\mathcal{U}^{(m_c)}$, denoted as $|\mathcal{U}^{(m_c)}|$, is equal to or greater than $U \cdot \Hat{t}_c$. In particular, this heuristic step, reported in Procedure~\ref{Alg.P1}, can be summarized as follows:
\begin{enumerate}[(i)]
\item Starting from the maximum MCS index $m_{\mathrm{MAX}}$ and $c = C$, we select the greatest MCS index such that the number of users in $\mathcal{U}^{(m_c)}$ is equal to or greater than $U \cdot \Hat{t}_c$.
\item Then, the index $c$ is decreased and the previous step is repeated by considering the MCS index range which goes from $m_c - 1$ to the minimum MCS index $m_{\mathrm{min}}$.
\item The procedure iterates while $\ell \geq 1$.
\end{enumerate}

The second step of the heuristic strategy aims at optimizing the variables $n^{(\ell,\ell)}$ (for $\ell = 1, \ldots, L$). In particular, let $\tilde{n}^{(\ell)}$ be the value of $n^{(\ell,\ell)}$ provided by the heuristic, where \mbox{$\tilde{\mathbf{n}} = \{\tilde{n}^{(t)}\}_{t = 1}^{L}$}. That optimization is summarized as follows:
\begin{enumerate}[(i)]
\item For any value of $\ell = 1, \ldots L$, $\tilde{n}^{(\ell)}$ is set equal to $k_\ell$ while $\tilde{n}^{(t)}$, for $t = \ell+1, \ldots, L$, is set to zero. Then the value of $\tilde{n}^{(\ell)}$ is progressively increased until $\mathrm{P}^{\mathrm{NOW}}_{1:\ell}(\tilde{\mathbf{n}}) \geq \Hat{P}$ does not hold and $\tilde{n}^{(\ell)} \leq \Hat{B}_\ell$.
\item The procedure iterates while $\ell \leq L$.
\end{enumerate}
It is straightforward to note that the aforementioned heuristic step requires a number of iterations which is equal to or less than $\sum_{t = 1}^{L}\left(\Hat{B}_t - k_t + 1\right)$.

\begin{algorithm}[tbd]
\floatname{algorithm}{Procedure}
\caption{Coded packet allocation for a NOW-RLNC service delivery using the MA pattern.}
\label{Alg.P2}
{\scriptsize \begin{algorithmic}[1]
\State $c \gets 1$
\State $\overline{n}^{(\ell,c)} \gets 0$ for any $\ell = 1, \ldots, L$ and $c = 1, \ldots, C$
\State $\overline{\mathbf{n}} = \{\overline{n}^{(\ell)}\}_{\ell = 1}^L$, where $\overline{n}^{(\ell)} \gets 0$ for any $\ell = 1, \ldots, L$

\For { $ l \gets 1, \ldots, L$ } \label{p2.forStart}
\While {$\mathrm{P}^{\mathrm{NOW}}_{1:\ell}(\overline{\mathbf{n}}) < \Hat{P}$ and $c \leq C$} \label{p2.whileStart}
	\State $\overline{n}^{(\ell,c)} \gets \overline{n}^{(\ell,c)} + 1$
    \State $\overline{n}^{(\ell)} \gets \sum_{t = 1}^{C}\overline{n}^{(\ell,t)}$ for any $\ell = 1, \ldots, L$
	\If { $\sum_{t = 1}^{L} \overline{n}^{(t,c)} = \Hat{B}_c$ } \label{p2.if1Start}
		\State $c \gets c + 1$
	\EndIf \label{p2.if1Stop}
\EndWhile \label{p2.whileStop}
\If { $\mathrm{P}^{\mathrm{NOW}}_{1:\ell}(\overline{\mathbf{n}}) < \Hat{P}$ and $c > C$ } \label{p2.if2Start}
	\State \textit{no solution can be found.}
\EndIf \label{p2.if2Stop}
\EndFor \label{p2.forStop}
\end{algorithmic}}
\end{algorithm}

Moving on to the MA pattern, to simplify our analysis, we impose that the number of subchannels has to be equal to the number of service layers, hence, $L = C$. However, the heuristic strategy we propose \emph{does not impose} that all the subchannels have to be used to deliver coded packets. This means that some subchannels could remain unassigned at the end of the allocation process. Concerning the first step of the heuristic strategy, we refer to the same procedure proposed for the SA pattern. For the second heuristic step, in this case, we refer to Procedure~\ref{Alg.P2}, which behaves as follows:
\begin{enumerate}[(i)]
\item We define $\overline{n}^{(\ell,c)}$, for $\ell = 1, \ldots, L$ and $c = 1, \ldots, C$, as the value of $n^{(\ell,c)}$ provided by the heuristic step. At the end of each iteration of the for-loop (lines~\ref{p2.forStart}-\ref{p2.forStop}), a set of values $\overline{n}^{(\ell,1)}, \overline{n}^{(\ell,2)}, \ldots, \overline{n}^{(\ell,C)}$ are derived, for every service layer. In particular, within the iteration associated with layer $\ell$, the value of $\overline{n}^{(\ell,c)}$ is incremented (\mbox{lines~\ref{p2.whileStart}-\ref{p2.whileStop}}) as long as the probability of recovering the first $\ell$ layers is smaller than $\Hat{P}$ and $\sum_{t = 1}^{L} \overline{n}^{(t,c)} \leq \Hat{B}_c$. If the $c$-th subchannel cannot hold more packets, the procedure switches to the next subchannel (lines~\ref{p2.if1Start}-\ref{p2.if1Stop}).
\item If the overall number of packets that can be conveyed by all the subchannels is not enough to deliver the coded packet stream associated with the first $\ell$ layers, the procedure cannot provide a valid allocation (lines~\ref{p2.if2Start}-\ref{p2.if2Stop}).
\end{enumerate}
It is straightforward to note that Procedure~\ref{Alg.P2} requires at most $\sum_{t = 1}^{C}\Hat{B}_t$ iterations.

Consider the second heuristic step of both SA and MA cases; both procedures generate \emph{the same} optimized number of coded packets associated to each service layer. The only difference between the two allocation patterns is that, in the second case, coded packets associated to the same service layer may be transmitted over multiple subchannels.

\subsection{Expanding Window Resource Allocation Strategy}\label{EWsec}
Similar to the NOW-RLNC case, we propose an optimization model suitable for the EW-based service delivery. Due to space limitations, we just focus on the MA allocation pattern.

Before giving the definition of the proposed EW-MA allocation model, it is worth recalling that, from the definition of the EW principle (see Section~\ref{subsec:EW_Rnc}), we know that user $u$ can recover the first $\ell$ service layers if the $\ell$-th window is recovered, or any window $t$, for $t = \ell+1, \ldots, L$, is recovered. Hence, we understand that user $u$ will recover the first $\ell$ service layers at least with probability $\Hat{P}$ if any of the windows $\ell, \ell+1, \ldots, L$ are recovered (at least) with probability $\Hat{P}$. For brevity, from~\eqref{eq.Pd_ew_complete}, we define the following indicator variable\footnote{In this paper, we refer to the logic expression $s_1 \vee s_2 \vee \cdots \vee s_T$ as $\bigvee_{t = \ell}^{T}s_t$, where $s_1, \ldots, s_T$ are logic statements.}
\begin{equation}
\mu_{u,\ell} = I\left(\bigvee_{t = \ell}^{L}\left\{\mathrm{P}^{\mathrm{EW}}_{1:t}(\mathbf{N}_u) \geq \Hat{P}\right\}\right).\label{eq.mu}
\end{equation}
In other words, $\mu_{u,\ell}$ is equal to one, if $u$ achieve a QoS level equal to or greater than $\ell$ with at least a probability of $\Hat{P}$.

The resource allocation model we propose, called EW-MA, can be expressed as follows:
\begin{align}
	\text{(EW-MA)} &  \quad  \mathop{\min_{m_1, \ldots, m_C}}_{N^{(1,c)}, \ldots, N^{(L,c)}} \,\,  \sum_{\ell=1}^{L} \sum_{c=1}^{C} N^{(\ell,c)} \label{EW-MA.of}\\
    \text{subject to} &   \quad \sum_{u = 1}^U \mu_{u,\ell} \geq U \,\, \Hat{t}_\ell \,\,\,\,\quad\quad\text{$\ell = 1, \ldots, L$}\label{EW-MA.c1}\\
                      &   \quad m_{c-1} < m_{c} \quad\quad\quad\quad\,\, \text{$c = 2, \ldots, L$}\label{EW-MA.c2}\\
                      &   \quad 0 \leq \sum_{\ell=1}^{L} N^{(\ell,c)} \leq \Hat{B}_c \quad  \text{$c = 1, \ldots, C$}\label{EW-MA.c3}
\end{align}
Also in this case, the objective function~\eqref{EW-MA.of} expresses the overall number of coded packet transmissions. Furthermore, constraint~\eqref{EW-MA.c1} imposes that the first $\ell$ service layers are recovered at least with probability $\Hat{P}$ by a fraction of users which shall not be smaller than $\Hat{t}_{\ell}$. Similarly to the NOW-SA and NOW-MA models, constraints~\eqref{EW-MA.c2} and~\eqref{EW-MA.c3} allow the model to exploit the heterogeneity of users.

Unfortunately, the EW-MA model is also a complex integer optimization problem, whose complexity is caused by the coupling constraints among optimization variables given by~\eqref{EW-MA.c1} and~\eqref{EW-MA.c3}. To this end, once again, we resort to a two-step heuristic strategy to find a good quality solution of EW-MA, in a finite number of steps.

Once more, for the first step, we refer to the same procedure adopted for the NOW-based allocation models. Let us define $\overline{N}^{(\ell,c)}$ as the value of $N^{(\ell,c)}$ provided by the heuristic step and $\overline{\mathbf{N}} = \{\overline{N}^{(\ell)}\}_{\ell = 1}^L$, where \mbox{$\overline{N}^{(\ell)} = \sum_{c=1}^{C}\overline{N}^{(\ell,c)}$}.  Starting from Procedure~\ref{Alg.P2}, the second heuristic step has been defined as follows:
\begin{enumerate}[(i)]
\item For $\ell = 1$ and $c = 1$, $\overline{N}^{(\ell,c)}$ is set to one, while $\overline{N}^{(\ell^\prime,c^\prime)} = 0$, for $\ell^\prime = \ell, \ldots, L$ and \mbox{$c ^\prime= c, \ldots, C$}. The value of $\overline{N}^{(\ell,c)}$ is gradually increased until \mbox{$\mathrm{P}^{\mathrm{EW}}_{1:\ell}(\overline{\mathbf{N}}) \geq \Hat{P}$} does not hold and $\sum_{t = 1}^{C}\overline{N}^{(\ell,t)} \leq \Hat{B}_t$. 
If the subchannel $c$ cannot hold more coded packets, coded packets will be gradually allocated on the next subchannel and the index $c$ is set equal to $c + 1$.
\item The value of the index $\ell$ is increased and the previous steps are repeated. The procedure iterates while $\ell \leq L$ and $c \leq C$.
\end{enumerate}
Finally, likewise to Procedure~\ref{Alg.P2}, the aforementioned heuristic step iterates for at most $\sum_{t = 1}^{C}\Hat{B}_t$ times.

\section{H.264/SVC Service Delivery over \mbox{LTE-Advanced} eMBMS Networks}\label{sec:eMBMS}
In order to give an overview of a possible practical implementation of the proposed standard-independent modelling and resource allocation strategies, we refer to the \mbox{LTE-A} standard. Since the first release of \mbox{LTE-A}, PtM communications are managed by means of the eMBMS framework~\cite{6353684}.

In the remaining part of the paper, we concentrate on a particular way of delivering PtM services, known as Single Cell-eMBMS (SC-eMBMS) transmission mode~\cite{sesia2011lte}. More precisely, we consider a network scenario formed by a base station, henceforth referred to as \emph{target base station}, which delivers a layered video service to a set of users forming a Multicast Group (MG), hereafter called \emph{target MG}. We also assume that all the multicast users are associated to the target base station. In addition, the target base station is surrounded by several interfering base stations, which impair service transmissions to the target MG.

\subsection{Network-coded Video Transmission over eMBMS Networks}\label{subsec.stack}
\begin{figure}[tbd]
\centering
\subfloat[MAC and physical layers.]{\label{fig.lteStack}
		\hspace{-2mm}\includegraphics[width=0.50\columnwidth]{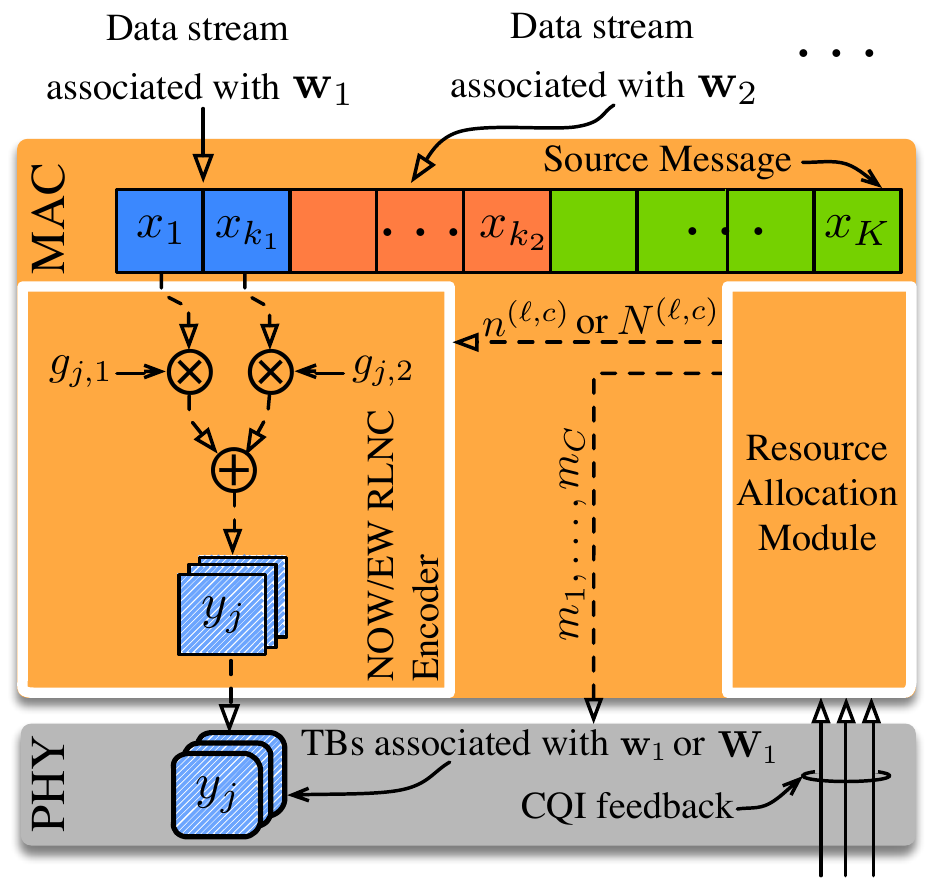}
}
\subfloat[Broadcast erasure subchannels.]{\label{fig.lteFrame}
		\includegraphics[width=0.47\columnwidth]{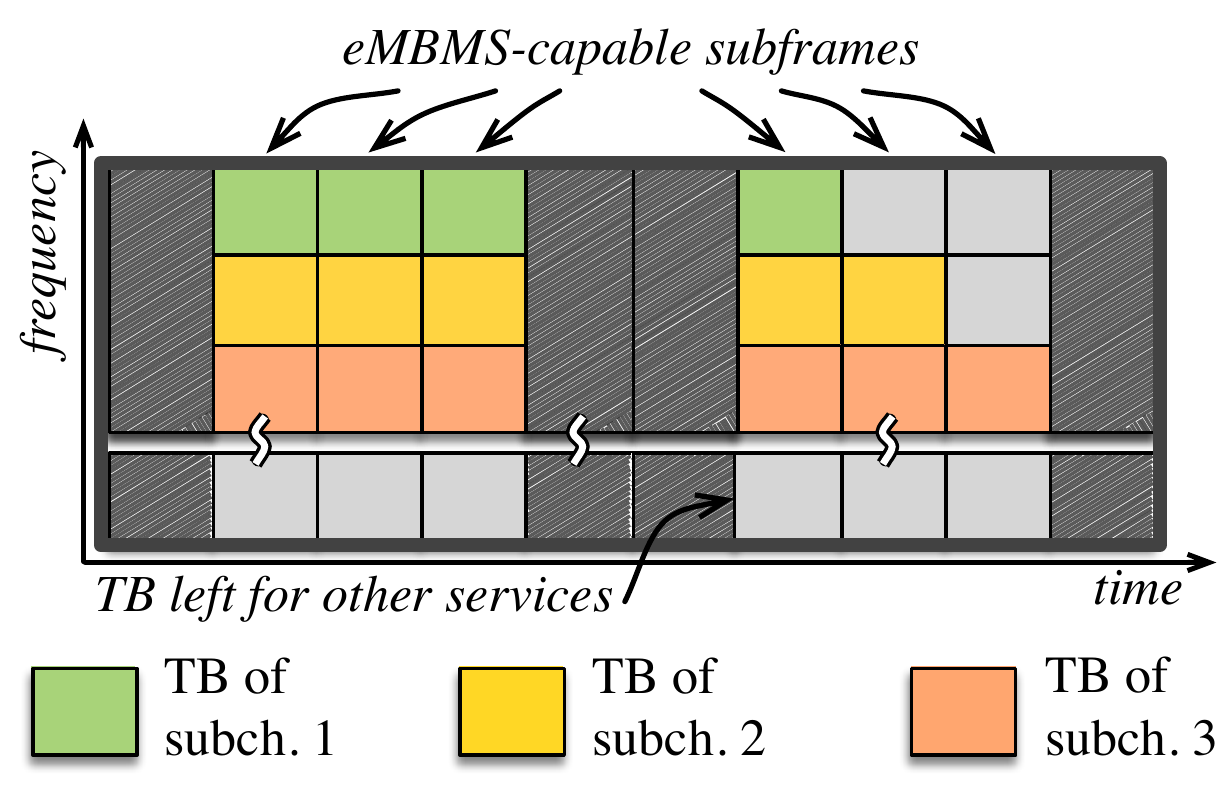}
}

\caption{A part of the considered \mbox{LTE-A} protocol stack and model of $C = 3$ broadcast erasure subchannels that span one \mbox{LTE-A} radio frame.}
\end{figure}

In our network scenario, the PtM multimedia service multicast by the target base station is a H.264/SVC video stream formed by $L$ different layers. In particular, the first layer, called \emph{base layer}, provides a basic reconstruction quality, which is gradually improved by the remaining $L-1$ layers, called \emph{enhancement layers}. In agreement with the layered message structure presented in Section~\ref{sec:system}, the level of the user QoS improves as the number of \emph{consecutive layers} (starting from the base layer) that can be successfully recovered increases.

We assume that each video layer is provided as an independent input of the \mbox{LTE-A} stack. More specifically, the data stream of each layer passes through the Packet Data Conversion Protocol and Radio Link Control layers then, it is forwarded to the Media Access Control (MAC) layer. Since each video layer has to be delivered by means of the NOW- or EW- RLNC approaches (see Section~\ref{subsec:NO_Rnc} and~\ref{subsec:EW_Rnc}), we refer to a modified MAC layer, similar to that proposed in~\cite{6353397}, which is in charge of all the network coding operations.

The layered video service produced by a H.264/SVC encoder can be modeled as a stream of Group of Pictures (GoPs)~\cite{h264}. Each GoP is characterized by fixed number of frames and has a fixed time duration $d_{\mathrm{GoP}}$. In particular, the value of $d_{\mathrm{GoP}}$ can be easily obtained by dividing the number of frames of a GoP with the video frame rate. Since the next GoP should be recovered (with a certain QoS) at least by the end of the currently reproduced one, the transmission time of each GoP shall not exceed $d_{\mathrm{GoP}}$.

Given that the decoding process of a H.264/SVC service is performed on a GoP-by-GoP basis, a GoP in our system model represents a layered source message to be delivered according to the network coding principle. We recall from Section~\ref{subsec:NO_Rnc} that $k_\ell$ is the number of source packets forming the $\ell$-th layer $\mathbf{w}_\ell$ of the source message. Consider Fig.~\ref{fig.lteStack}, the MAC layer segments the data stream, forwarded by the higher protocol layers and associated with the $\ell$-th video layer of a GoP, into $k_\ell$ source packets with the same bit length $H$. Let $\nu_\ell$ be the bitrate associated with the $\ell$-th video layer observed at the MAC layer. The term $k_\ell$ can be defined as $k_\ell = \left\lceil \left(\nu_\ell \cdot d_{\mathrm{GoP}}\right) / H\right\rceil$.

For each GoP, the MAC layer produces streams of coded packets, each of which has the same size of a source packet. In particular, in the case of the NOW-RLNC, the MAC layer produces one stream of coded packets per video layer. On the other hand, in the case of the EW-RLNC case, one stream per window is generated. We assume that the selection process of coding coefficients is initialized by random number generator (RNG) seeds that are delivered to the multicast user as part of \mbox{LTE-A} signalling data. Delivered RNG seeds are used to re-generate coding coefficients~\cite{6353397}.

Each coded packet is forwarded to the physical layer and mapped onto one Transport Block (TB). A TB is a frequency-time structure characterized by a fixed transmission time duration $d_{\mathrm{TTI}} = \SI{1}{\milli\second}$ equal to one Transmission Time Interval (TTI). Each TB may consist of one or more ``resource block pairs'', which are frequency-time resource units that span a bandwidth of \SI{180}{\kilo\hertz} and have the same transmission time duration of a TB. In other words, the TB bandwidth is an integer multiple of \SI{180}{\kilo\hertz}. Furthermore, a TB is transmitted with a certain MCS~\cite{sesia2011lte}.

We remark that the actual number of bits (referred to as \emph{bit capacity}, in this paper) that a resource block pair can hold depends on the MCS in use. Assuming that a TB can hold just one coded packet, both the number of resource block pairs per TB and the source/coded packet size $H$ have to be selected in order to fit, as tightly as possible, the bit capacity of a TB. To this end, let $\mathrm{N}_{\mathrm{B},m}$ and $\mathrm{N}_{\mathrm{C},m}$ be the number of resource block pairs forming a TB and the bit capacity of a resource block pair, for the $m$-th MCS, respectively. In this paper, $H$ and $\mathrm{N}_{\mathrm{B},m}$ values  have been obtained by solving the following min-max problem:
\begin{align}
 &  \quad  \mathop{\min\max}_{\mathrm{N}_{\mathrm{B},\mathrm{min}}, \ldots, \mathrm{N}_{\mathrm{B},\mathrm{MAX}}, H} \,\,  \mathrm{N}_{\mathrm{B},m} \mathrm{N}_{\mathrm{C},m} - H \label{MM.of}\\
    \text{subject to} &   \quad\mathrm{N}_{\mathrm{B},m} \mathrm{N}_{\mathrm{C},m} \geq H \quad\,\text{$m = m_{\mathrm{min}}, \ldots, m_{\mathrm{MAX}}$}\label{MM.c1}\\
                      &   \quad\mathrm{N}_{\mathrm{B},m} \leq \Hat{\mathrm{N}}_{\mathrm{B}} \quad\,\,\,\,\,\,\,\,\,\,\,\, \text{$m = m_{\mathrm{min}}, \ldots, m_{\mathrm{MAX}}$}\label{MM.c2}
\end{align}
where the objective function~\eqref{MM.of} minimizes the maximum unused bit capacity per TB, for all the possible MCSs. Constraint~\eqref{MM.c1} ensures that the TB bit capacity is at least equal to $H$, for any MCSs. In addition, constraint~\eqref{MM.c2} imposes that the number of resource block pairs per TB does not exceed a maximum value equal to $\Hat{\mathrm{N}}_{\mathrm{B}}$. Note that~\eqref{MM.of}-\eqref{MM.c2} is an integer optimization problem but it has a modest complexity and can be solved by means of a basic branch-and-bound strategy~\cite{Couenne}.

\subsection{MAC Layer Augmented Resource Allocation Capabilities}\label{subsec.RA}
Even though the eMBMS framework enables \mbox{LTE-A} to manage PtM service transmission, the standard delegates the definition and implementation of all the resource allocation operations to the manufactures. However, the standard imposes that the MAC layer is in charge of all the scheduling and resource allocation tasks~\cite{TR_36_321}. For these reasons, we assume that the considered network coding-capable MAC layer is also in charge of allocating resources according to the resource allocation strategies presented in Section~\ref{subsec:raModels}. To this end, we update the subchannel definition given in Section~\ref{sec:system}.

Consider Fig.~\ref{fig.lteFrame}, which shows the structure of one \mbox{LTE-A} radio frame. One frame is composed of $10$ subframes, each subframe has a transmission time duration equal to $1$ TTI. At most $6$ out of $10$ subframes of a radio frame can be used to deliver eMBMS traffic~\cite{sesia2011lte}, while the remaining subframes are dedicated to PtP traffic. Consider subchannel $c$, we remark that the maximum number $\Hat{B}_c$ of coded packets that can be transmitted over it, during a given time interval, is fixed. Since a TB can hold just one coded packet, we define the subchannel $c$, as shown in Fig.~\ref{fig.lteFrame}, as a group of $\Hat{B}_c$ TBs, transmitted over eMBMS-capable subframes. In particular, we impose that just one TB per-subchannel can be delivered during a TTI.

For simplicity, in the considered \mbox{LTE-A} scenario, we assumed that $\Hat{B}_c = \Hat{B}$ (for $c = 1, \ldots, C$), and that the considered fraction of eMBMS-capable subframes per radio frame is $0.6$, i.e., $6$ out of $10$ subframes. Due to the fact that each GoP shall be delivered before transmission of the next GoP begins, the value of $\Hat{B}_c$ shall not be greater than $\Hat{d}_{\mathrm{GoP}} = \left\lfloor 0.6 \cdot \left(d_{\mathrm{GoP}} / d_{\mathrm{TTI}}\right)\right\rfloor$ TTIs. 

In \mbox{LTE-A} systems, the reception of TB, which adopts a given MCS, is acceptable as long as the TB error rate experienced by a user $u$ is equal to or smaller than $0.1$~\cite{sesia2011lte}. The standard allows users to provide Channel Quality Indicator (CQI) feedback to the base station about their propagation conditions. In particular, the CQI feedback provided by a user $u$ indicates the greatest MCS index $m \in [1, 15]$ (see \mbox{Table~7.2.3-1~\cite{TR_36_213}}) such that the TB error probability of $u$ is equal to or smaller than $0.1$~\cite{sesia2011lte}. To this end, we set $\Hat{p} = 0.1$ in~\eqref{eq.CQI}. Obviously, the actual PER experienced by each user of the target MG is unknown to the target base station. However, as reported in Fig.~\ref{fig.lteStack}, the \mbox{LTE-A} standard imposes that CQI feedback are directly forwarded to the MAC layer. Hence, it is reasonable to assume that the proposed resource allocation strategies can easily access the CQI information. Owing to the lack of knowledge of the user PER, the target base station approximates the user PER as $p_u(m_c) = \Hat{p}$ if $M(u) \geq m_c$, otherwise $p_u(m_c) = 1$. As a consequence, the definition of $p_u$, provided by~\eqref{eq.p}, is updated as follows:
\begin{equation}
p_u \cong \!\!\max_{c = 1, \ldots, C}\!\left(\Hat{p} \,\,\Big|\,\, M(u) \geq m_c \wedge \sum_{\ell = 1}^{L} n^{(\ell,c)} > 0\right)\! \label{eq.pLTE}
\end{equation}
where $M(u)$ is equal to the MCS index reported in the CQI feedback provided by user $u$. For the sake of clarity, we note that the approximation of $p_u$, given in~\eqref{eq.pLTE}, is considered only by the target base station during the resource allocation operations. On the other hand, all the analytical results and performance assessment, presented in the following sections, will refer to the user PER expression provided in~\eqref{eq.p}.

Consider again Fig.~\ref{fig.lteStack}, all the resource allocation operations can be ideally modelled as a functional block of the MAC layer. In the case of the proposed resource allocation strategies, the resource allocation module provides the optimized $n^{(\ell,c)}$ or $N^{(\ell,c)}$, for $\ell = 1, \ldots, L$ and $c = 1, \ldots, C$, to the network coding encoder. In addition, the optimized MCS values $m_1, \ldots, m_C$, associated to each subchannel, are forwarded to the physical layer, which is in charge of transmitting each TB.

Even though this section considered the LTE-A standard, we point out what follows: (i) The generic modelling of Sections~\ref{sec:system} and~\ref{subsec:raModels} can be easily adapted to any OFDMA-based system able to manage PtM communications and hence also future LTE-A releases, (ii) The considered RLNC schemes and the proposed resource allocation strategies should be plugged into the protocol stack layer in charge of allocating radio resources and, (iii) Our practical implementation proposal can be easily adapted to any kind of multimedia layered service.

\section{Analytical Results}\label{sec:results}
We investigate the performance of the proposed resource allocation strategies by considering an \mbox{LTE-A} network formed by a $19$ macro-base stations. In particular, we assume that the cell controlled by the target base station (hereafter called \emph{target cell}) is surrounded by $18$ interfering macro-base stations, organized in two concentric rings. Each base station manages three hexagonal sectors per cell. Concerning the physical layer and transmission parameters, we referred to the 3GPP's benchmark simulation scenario, called \emph{Case 1} scenario~\cite{TR_36_814}, where base stations are characterized by an inter-site-distance of \SI{500}{\meter}. Furthermore, we assumed that users forming the target MG are placed outdoors. Hence, all the physical layer parameters have been set by following the guidelines provided in \mbox{Tables~A.2.1.1-2} and~\mbox{A.2.1.1.2-3} of~\cite{TR_36_814}. The first part of Table~\ref{tab.params} summarizes all the remaining system parameters we considered.

In order to provide an effective user QoS assessment, we considered a user distribution characterized by a high heterogeneity from the point of view of the experienced propagation conditions. This means that each user is characterized by a different Signal to Interference plus Noise Ratio (SINR) and hence, a different PER. In particular, we refer to a target MG of $U=80$ users that are placed along the radial line representing the symmetry axis of one sector of the target cell. The first user is placed at a distance of \SI{90}{\meter} from the target base station and the distance between two consecutive users is \SI{2}{\meter}.

\begin{table}[tbp]
\centering
\caption{Main simulation parameters.}
\label{tab.params}
{\scriptsize\begin{tabular}{cc|c|c|}
\cline{3-4}  \multicolumn{2}{c|}{} & \textbf{Parameter} 	& 	\textbf{Value}						\\
\hline \multicolumn{2}{|c|}{\multirow{8}{*}{\rotatebox[origin=c]{90}{\textbf{Physical Layer}}}} & Inter-Site-Distance		& 	500 m								\\
\cline{3-4} \multicolumn{2}{|c|}{} & System Bandwidth		& 	20 MHz								\\
\cline{3-4} \multicolumn{2}{|c|}{} & Transmission Scheme			& 	SISO									\\
\cline{3-4} \multicolumn{2}{|c|}{} & Duplexing Mode				& 	FDD									\\
\cline{3-4} \multicolumn{2}{|c|}{} & Carrier Frequency				& 	2 GHz								\\							
\cline{3-4} \multicolumn{2}{|c|}{} & Transmission Power				& 	46 dBm per sector					\\
\cline{3-4} \multicolumn{2}{|c|}{} & Base Station and User  Antenna Gains		& 	see Table A.2.1.1-2~\cite{TR_36_814}	\\
\cline{3-4} \multicolumn{2}{|c|}{} & Pathloss	and Penetration Loss & 	see Table A.2.1.1.2-3~\cite{TR_36_814} \\
\cline{3-4} \multicolumn{2}{|c|}{} & Channel Model & 	ITU-T PedA~\cite{Access2013} \\
\cline{1-4} \multicolumn{2}{|c|}{\multirow{9}{*}{\rotatebox[origin=c]{90}{\textbf{MAC Layer}}}} & $C$ &  $3$ subchannels\\
\cline{3-4} \multicolumn{2}{|c|}{} & $\Hat{\mathrm{N}}_{\mathrm{B}}$ &  $6$ resource block pairs\\
\cline{3-4} \multicolumn{2}{|c|}{} & $\mathrm{N}_{\mathrm{C},m}$, for $m = 4, \ldots, 15$ &  see Table 10.1~\cite{sesia2011lte}\\
\cline{3-4} \multicolumn{2}{|c|}{} & $\mathrm{N}_{\mathrm{B},m}$, for $m = 4, \ldots, 15$ &  $6, 4, 3, 3, 2, 2, 2, 1, 1, 1, 1, 1$\\
\cline{3-4} \multicolumn{2}{|c|}{} & $H$ &  $32$ KB\\
\cline{3-4} \multicolumn{2}{|c|}{} & $\Hat{P}$ &  $0.99$\\
\cline{3-4} \multicolumn{2}{|c|}{} & $\Hat{t}_\ell$, for $\ell = 1, \ldots L$ &  $0.99, 0.8, 0.6$\\
\cline{3-4} \multicolumn{2}{|c|}{} & $\Hat{B}_c$, for $c = 1, \ldots, C$ &  $K + \lceil K/2 \rceil$ TB transmissions\\
\cline{3-4} \multicolumn{2}{|c|}{} & $q$ &  $[2, 2^8]$\\
\hline
\end{tabular}}
\end{table}

In this performance investigation, we refer to two different video streams encoded using the H.264/SVC Coarse Grain Scalability (CGS) principle. Each layer of a CGS stream successively increases the fidelity of any video frame. In order to do so, H.264/SVC CGS adopts those forms of spatial scalability such that the combination of one or more consecutive layers gives the same spatial frame resolution~\cite{6194978}. Both video streams belong to the video trace database, provided as a companion of~\cite{6025326}, and developed for network performance evaluation purposes. The first stream is the \emph{News CIF} ($352\!\times\!288$) video sequence~\cite{NewsCIF} composed by $L = 3$ layers, with GoPs of size $16$ frames and video frame rate of $30$ frame-per-second (fps). The second stream is the \emph{Blue Planet} ($1920\!\times\!1088$) video sequence~\cite{BluePlanet} that consists of $L = 3$ layers, GoPs of size $16$ frames and video frame rate of $24$ fps.

It is worth noting that the bit rate of the video stream obtained by combining all the layers of Blue Planet is \mbox{$2.8$-times} greater than that of News CIF video stream. In addition to the main characteristics of the considered video streams, Table~\ref{tab.streams} gives the maximum bitrate $\nu_1\, \ldots, \nu_L$ per-video layer, for each stream. Furthermore, as a performance metric of the video fidelity, Table~\ref{tab.streams} provides also the average Peak Signal-to-Noise Ratio (PSNR) $\rho_\ell$ achieved after successfully recovering the first $\ell$ video layers, for $\ell = 1, \ldots, L$~\cite{6025326}.

In order to inspect the impact of different resource allocation models on the target MG, we developed a system level MATLAB simulator. In particular, we refer to the simulation framework proposed in~\cite{6353397}. Hence, given the physical layer parameters of Table~\ref{tab.params} and Eq.~(1) of~\cite{6353397}, we evaluated the average SINR value associated to each user in the target MG. Consider~\eqref{eq.p}, in order to assess the user performance, we need the PER value $p_u(m_c)$ associated to the user $u$ and MCS $m_c$. Unlike~\cite{6353397}, we relied on the \mbox{LTE-A} downlink link level simulator presented in~\cite{Access2013} to obtain the value of $p_u(m_c)$, as a function of the average user SINR. In particular, for any average SINR value, $p_u(m_c)$ is set equal to the PER value obtained from the \mbox{LTE-A} downlink link level simulator and averaged over $10^4$ simulation runs. Since we are concerned with stationary and low-mobility users, link level simulations have been performed by considering the ITU-T PedA channel model~\cite{Access2013}. Hence, by using~\eqref{eq.CQI}, it is straightforward to emulate the CQI feedback that users provide to the target base station (see Section~\ref{subsec.RA}).

We remark that the MCS index advertised by CQI feedback may span the interval $[1, 15]$. Since the bitrates ensured by MCSs $1$-$3$ are too small\footnote{For a TB formed by one resource block pair, MCS index $m=3$ ensures a bitrate smaller than $26.7$ kbps, at net of all the signalling information.} compared to bitrates of the considered video streams, users providing CQI feedback with MCS indexes less than $4$ are excluded from the optimization process. For this reason, we set $m_{\mathrm{min}}$ equal to $4$, while $m_{\mathrm{MAX}}$ is kept equal to $15$.

\begin{table}[tbp]
\centering
\caption{H.264/SVC video streams considered.}
\label{tab.streams}
{\scriptsize\begin{tabular}{|c|c|c|c|c|}
\hline \multirow{2}{*}{\textbf{Stream}} & \hspace*{0mm}{No. Frames} & Frame Rate & \hspace{-1.5mm}$\nu_1, \ldots, \nu_L$\hspace{-1.5mm} & \hspace*{0mm}$\rho_1, \ldots \rho_L$\\
 & \hspace*{0mm}{per GoP} & (fps) & (Mbps) & (dB) \\
\hline News CIF & \multirow{3}{*}{$16$} & \hspace*{0mm}{\multirow{3}{*}{$30$}} & \multirow{3}{*}{\hspace{-1.5mm}$2.45, 2.45, 7.35$\hspace{-1.5mm}} & \hspace*{-1.5mm}\multirow{3}{*}{$31.6, 37.4, 43.7$}\hspace*{-1.5mm}\\
$L = 3$ &  & &  & \\
\hspace*{-1.5mm}$\Hat{d}_{GoP} = 320$\hspace*{-1.5mm} &  & &  & \\
\hline Blue Planet & \multirow{3}{*}{$16$} & \hspace*{0mm}{\multirow{3}{*}{$24$}} & \multirow{3}{*}{\hspace{-1.5mm}$1.96, 2.94, 19.60$\hspace{-1.5mm}} & \hspace*{-1.5mm}\multirow{3}{*}{$36.6, 44.5, 51.2$}\hspace*{-1.5mm}\\
 $L = 3$ &  &  &  & \\
 \hspace*{-1.5mm}$\Hat{d}_{GoP} = 320$\hspace*{-1.5mm} &  & &  & \\
\hline
\end{tabular}}
\end{table}

Each video layer of a video stream is delivered by the target base station over $C = 3$ subchannels, as described in Sections~\ref{sec:system} and~\ref{subsec:raModels}. As noted in Section~\ref{sec:eMBMS}, the number $\mathrm{N}_{\mathrm{B},m}$ of resource block pairs forming a TB depends on the MCS index $m$ used to transmit it. Assuming that each TB cannot consists of more than $\Hat{\mathrm{N}}_{\mathrm{B}} = 6$, the solution to problem~\eqref{MM.of}-\eqref{MM.c2} is reported in Table~\ref{tab.params}. We remark also that the source/coded packet bit size $H$ is part of the aforementioned solution.

Consider the remaining MAC layer simulation parameters of Table~\ref{tab.params}, they are related to the resource allocation strategies (see Section~\ref{subsec:raModels}). In particular, we assumed that consecutive video layers, starting from the base layer, shall be recovered with at least a probability of $\Hat{P} = 0.99$. Furthermore, we imposed that at least $99$\% and $60$\% of the users forming the target MG shall experience the basic or the maximum QoS, respectively. For simplicity we assume that any subchannel consists of the same number of TBs. Having in mind that the transmission time duration of any layer of a GoP shall not be greater than $\Hat{d}_{\mathrm{GoP}}$ we set $\Hat{B}_c$ equal to $K + \lceil K/2 \rceil$, as a case of study.

\subsection{Performance Metrics and Benchmark}
Performance has been evaluated in terms of the total number of TB transmissions $\tau$ needed to deliver all video layers of a GoP. In the remaining part of the paper, we will refer to $\tau$ as the \emph{resource footprint}. From the expressions of the objective functions~\eqref{NO-SA.of} and~\eqref{EW-MA.of}, $\tau$ can be defined as follows:
\begin{equation}
\tau = \begin{cases}
			\displaystyle\sum_{\ell=1}^{L} \sum_{c=1}^{C} n^{(\ell,c)}\text{,} \quad \,\,\text{for NOW-RLNC}\\
			\displaystyle\sum_{\ell=1}^{L} \sum_{c=1}^{C} N^{(\ell,c)}\text{,} \quad \text{for EW-RLNC}
		 \end{cases}\label{eq.sigma}
\end{equation}
where the values of $n^{(\ell,c)}$ and $N^{(\ell,c)}$ have been optimized by the resource allocation strategies presented in Section~\ref{subsec:raModels}. From~\eqref{eq.noFC} or~\eqref{eq.Pd_ew_complete} we also evaluated user performance in terms of the probability that a user $u$ recovers the first $\ell$ video layers. Furthermore, we considered, as a third performance metric, the maximum PSNR that user $u$ can achieve, defined as:
\begin{equation}
\rho(u) \!=\! \begin{cases}
			\displaystyle \max_{\ell = 1, \ldots, L}\left\{\rho_\ell \,\, \mathrm{P}^{\mathrm{NOW}}_{1:\ell}(\mathbf{n}_u)\right\}\text{,} \,\,\, \text{for NOW-RLNC}\\
			\displaystyle \max_{\ell = 1, \ldots, L}\left\{\rho_\ell \,\, \mathrm{P}^{\mathrm{EW}}_{1:\ell}(\mathbf{N}_u)\right\}\text{,} \,\,\,\,\, \text{for EW-RLNC.}
		 \end{cases}\!\!\!\!\label{eq.rho}
\end{equation}
Since the users of the target MG are regularly placed on the symmetry axis of the cell-sector, the value of $\rho(u)$ can be equivalently expressed in terms of the distance between the user $u$ and the target base station. In a similar way, parameter $\Hat{t}_\ell$ can be interpreted as the minimum distance, from the centre of the target cell, where a user shall recover the first $\ell$ video layers with a probability of at least $\Hat{P}$.

We provide performance comparisons among solutions of NOW-SA, NOW-MA and EW-MA, obtained by the proposed heuristic strategies and by directly solving the aforementioned problems using a genetic strategy (we refer to this kind of solutions as \emph{direct solutions})~\cite{Deep2009505}. Even though, the direct solution can be considered as a good approximation of the optimal solution of the proposed problems, it is worth noting that a genetic strategy cannot be considered a viable alternative to solve the proposed optimization models in a practical scenario because of its computational complexity~\cite{goldberg2013genetic}.

Both the direct and the heuristic solutions of the proposed resource allocation strategies have been compared with a MrT transmission strategy that relies on a standard \mbox{LTE-A} protocol stack. In other words, we referred to a protocol stack which does not adopt RLNC-based service multicasting and does not rely on any AL-FEC strategy. For the implementation of the considered MrT strategy, we refer to the resource allocation strategy proposed in~\cite{4917957,5452675} which aims at maximizing the sum of the video quality experienced by each user. In particular, this goal is achieved by optimizing the MCS index $m_\ell$ used to deliver the TB stream holding data associated with the $\ell$-th video layer, for $\ell = 1, \ldots, L$. 

It is worth noting that both~\cite{4917957} and~\cite{5452675} implicitly refer to a concept that is similar to the SA pattern. Specifically, data streams associated to different video layers are independently transmitted to the target MG. Assume that the $\ell$-th video layer is delivered with the MCS with index $m_\ell$. We understand that, in the case that the target base station relies on the standard \mbox{LTE-A} protocol stack, the uncoded transmission of TBs associated to video stream $\ell$ is equivalent to the transmission of all the $k_\ell$ TBs defining the $\ell$-th layer $\mathbf{w}_\ell$ of a GoP. In order to make fair comparisons, we referred here to the same values of $\mathrm{N}_{\mathrm{B},m}$ reported in Table~\ref{tab.params}. For these reasons, the probability $\mathrm{P}^{\mathrm{MrT}}_{1:\ell}$ that user $u$ recovers the first $\ell$ layers can be expressed as $\mathrm{P}^{\mathrm{MrT}}_{1:\ell} = \mathrm{P}^{\mathrm{MrT}}_{1:\ell-1} \cdot [1-p_{u}(m_\ell)]^{k_\ell}$ where, $\mathrm{P}^{\mathrm{MrT}}_{1:1} = [1-p_{u}(m_1)]^{k_1}$. In this case, the maximum PSNR that $u$ can achieve is $\rho(u) = \displaystyle \max_{\ell = 1, \ldots, L}\left\{\rho_\ell \cdot \mathrm{P}^{\mathrm{MrT}}_{1:\ell}\right\}$. Hence, we expressed the considered MrT strategy as follows:
\begin{align}
	\text{(MrT)} &  \quad  \max_{m_1, \ldots, m_L} \sum_{u=1}^{U} \rho(u) \label{MrT.of}\\
    \text{subject to} &   \quad m_{\ell-1} < m_{\ell} \,\,\,\,\,\quad\quad\text{$\ell = 2, \ldots, L$.}\label{MrT.c1}
\end{align}
As well as in the case of the proposed resource allocation strategies, the exact value of $p_{u}(m_\ell)$ is unknown at the target base station side. Hence, during the resource allocation based on MrT, the PER expression is approximated as $p_{u}(m_\ell) \cong \Hat{p}$ if $M(u) \geq m_\ell$, otherwise $p_{u}(m_\ell) \cong 1$.

\begin{figure}[tbd]
\centering
\includegraphics[width=0.99\columnwidth]{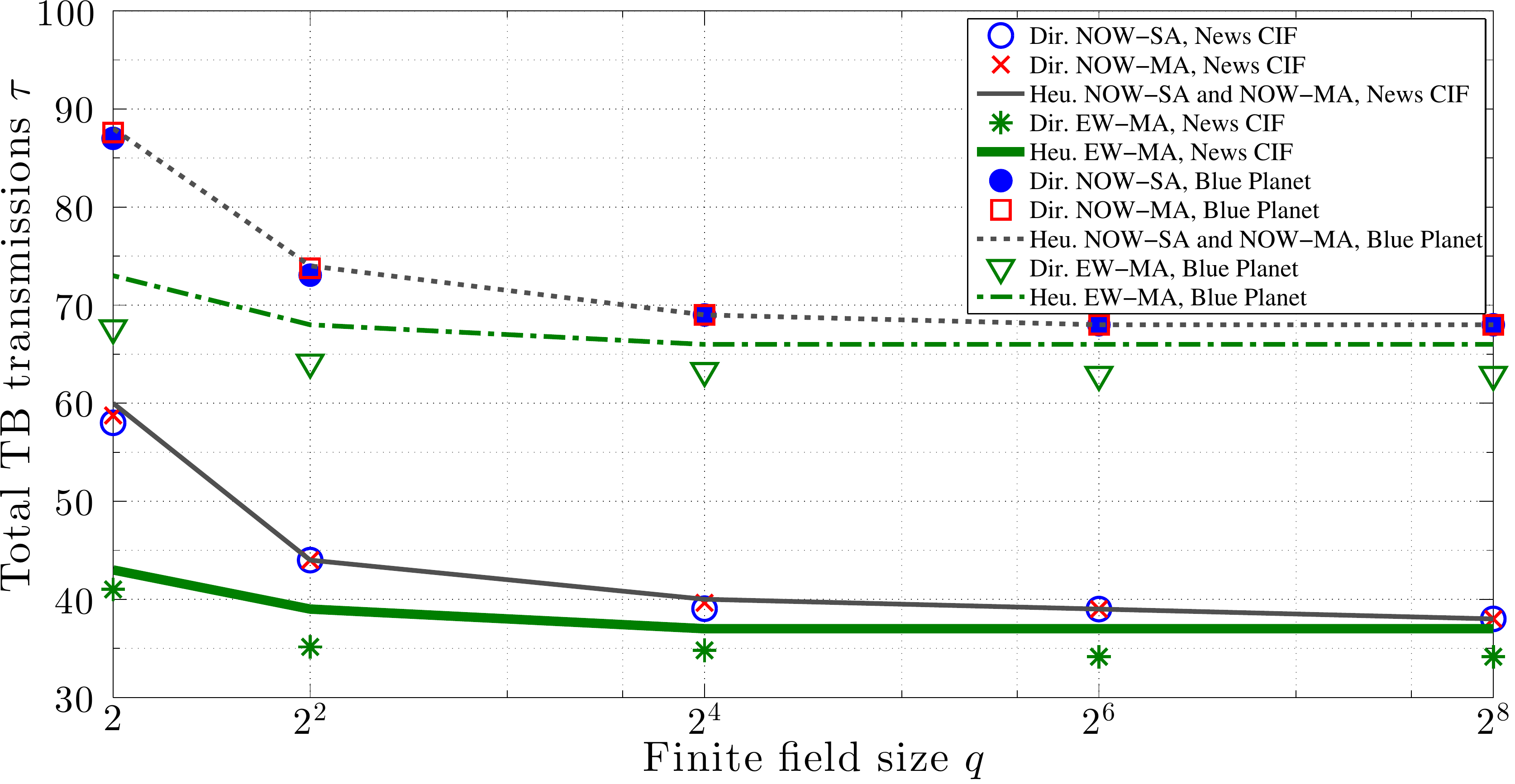}
\caption{Overall number of TB transmissions associated with all the proposed resource allocation frameworks.}
\label{fig.footprint}
\end{figure}

\subsection{Assessment of the Heuristic Solutions}
Let us start our performance investigation from Fig.~\ref{fig.footprint}, it compares the number of TB transmissions, represented by $\tau$, which are associated with the direct (``Dir.'') and heuristic (``Heu.'') solutions, of all the proposed resource allocation strategies, as a function of the finite field size $q$ over which all the RLNC-related operations are performed. The figure shows results for both News CIF and Blue Planet streams. Due to the fact that $\tau$ represents the value of the objective functions of the proposed optimization models, it allows us to inspect the performance gap between each proposed heuristic strategy and the corresponding direct solution. We remark that the number of TB transmissions associated with a direct solution is unlikely to be greater than that associated with a heuristic solution~\cite{Deep2009505}. As clearly shown in Fig.~\ref{fig.footprint}, the performance gap between the heuristic and the direct solutions is negligible. In particular, the gap is at most equal to $2$, $1$ and $5$ TBs for the NOW-SA, NOW-MA and EW-MA models, respectively. For this reason, in the rest of this section, we refer only to the heuristic solutions of the proposed resource allocation models.

We also observe in Fig.~\ref{fig.footprint} that the value of $\tau$, of any resource allocation model, decreases as the the value of $q$ increases. We understand that, for an increasing value of $q$, the probability of receiving coded packets that are linearly dependent with the previous ones decreases. As a consequence, the resource footprint of each allocation strategy decreases, as the finite field size increases. However, for small finite field sizes, there is a remarkable gap between any solution based on a \mbox{NOW-RLNC} strategy and the direct/heuristic EW-MA solution. In particular, for $q = 2$, the gap between the heuristic solution of EW-MA and, either NOW-SA or NOW-MA, is equal to $17$ TBs.

\begin{figure}[tbd]
\centering
\includegraphics[width=0.99\columnwidth]{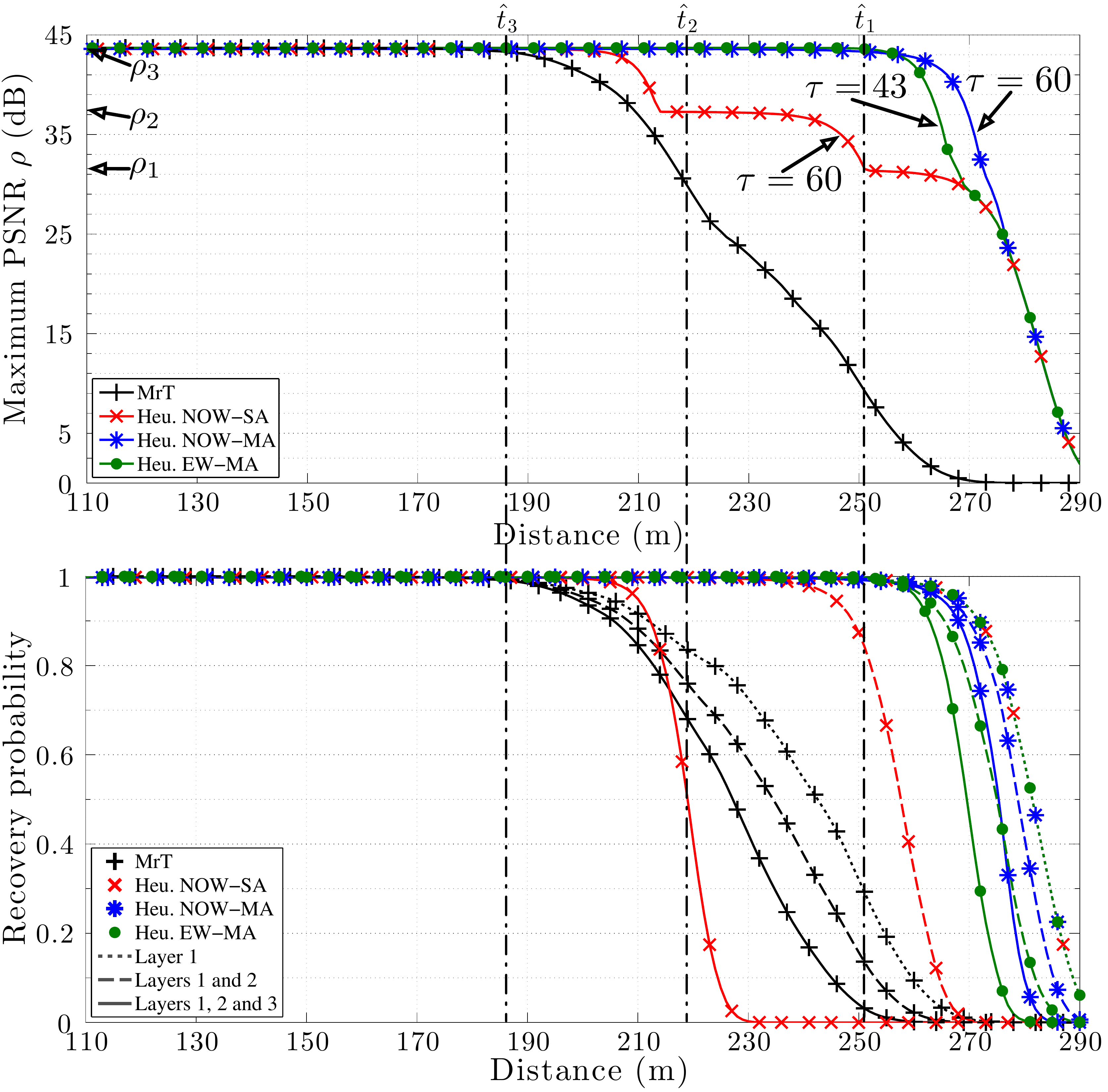}
\caption{Maximum PSNR and probability of recovering a given set of video layers associated with stream News CIF, for $q=2$.}
\label{fig.A_2}
\end{figure}

\subsection{Performance Evaluation of the Proposed Frameworks}\label{subsec.perfFr}
Focusing on a finite field with $q = 2$, Fig.~\ref{fig.A_2} shows both the maximum PSNR $\rho$, and the probability of recovering the first $\ell$ video layers as a function of the distance from the centre of the cell. We recall that the MrT strategy does not rely on any AL-FEC or RLNC-based strategy. In addition, due to the fact that the MrT aims at maximizing the sum of the video quality achieved by all the users, its performance in terms of coverage diverges from both that of the proposed strategies and the target performace. In particular, we note that the base video layer can be received at least with a probability of $0.99$ up to a distance of \SI{188}{\meter} from the centre of the cell. The MrT performance confirms the idea underlying the proposed optimization strategies; that is defining allocation models where the constraint set ensures that a target coverage and objective function minimizes the amount of resources needed to deliver the multicast service.

We see also in Fig.~\ref{fig.A_2} that, even though all the proposed allocation models meet the coverage constraints, strategies based on the MA pattern provide better coverage than that associated with the SA pattern. In particular, due to the fact that MA pattern can exploit the user heterogeneity better than the SA one, both NOW-MA and EW-MA can successfully deliver all the video layers up to a distance of \SI{252}{\meter}. On the other hand, the NOW-SA model ensures the maximum service quality only up to \SI{203}{\meter}. Furthermore, from Fig.~\ref{fig.A_2}, we understand that the $\tau$ value of the heuristic EW-MA strategy is $\sim 28$\% smaller than that of the NOW-MA and NOW-SA heuristic solutions. In particular, we can argue that both the NOW-MA and EW-MA strategies achieve almost the same coverage performance but the second one requires a smaller resource footprint. Finally, as expected (see Section~\ref{subsec:raModels}), both the heuristic NOW-SA and NOW-MA models are characterised by the same values of $\tau$.

\begin{figure}[tbd]
\centering
\includegraphics[width=0.99\columnwidth]{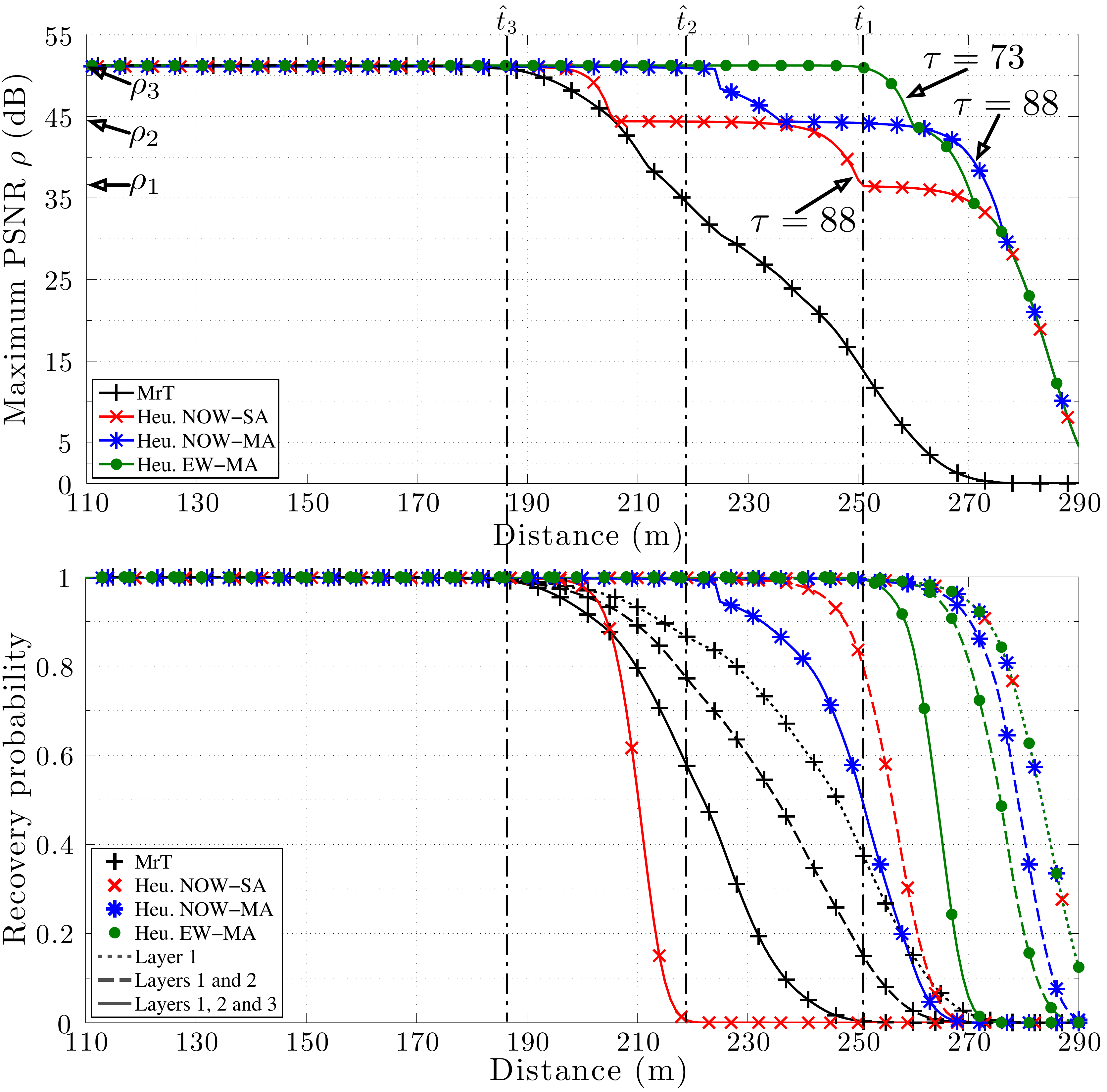}
\caption{Maximum PSNR and probability of recovering a given set of video layers associated with stream Blue Planet, for $q=2$.}
\label{fig.B_2}
\end{figure}

Fig.~\ref{fig.B_2} compares the same performance metrics considered in Fig.~\ref{fig.A_2} (for $q = 2$), associated with the stream Blue Planet. We remark, the overall bitrate of stream Blue Planet is greater than that of the stream News CIF. Also in this case, we note that all the proposed resource allocation solutions meet the target service constraints. As shown by Fig.~\ref{fig.footprint}, the $\tau$ value associated with the heuristic EW-MA strategy is $\sim 17$\% smaller than that of the heuristic NOW-SA/NOW-MA solution. Furthermore, the EW-MA strategy provides a resource allocation solution such that all the video layers can be successfully recovered up to a distance of \SI{252}{\meter}, which is \SI{28}{\meter} greater by than that ensured by NOW-MA. In accordance with stream News CIF, we observe that the heuristic NOW-SA provides allocation solutions such that all the video layers can be recovered up to a distance that is \SI{27}{\meter} (\SI{55}{\meter}) smaller, respectively, than that associated with the heuristic NOW-MA (EW-MA). Finally, also in this case, the performance of MrT diverges from the performance of the proposed strategies.

\begin{figure}[tbd]
\centering
\includegraphics[width=0.99\columnwidth]{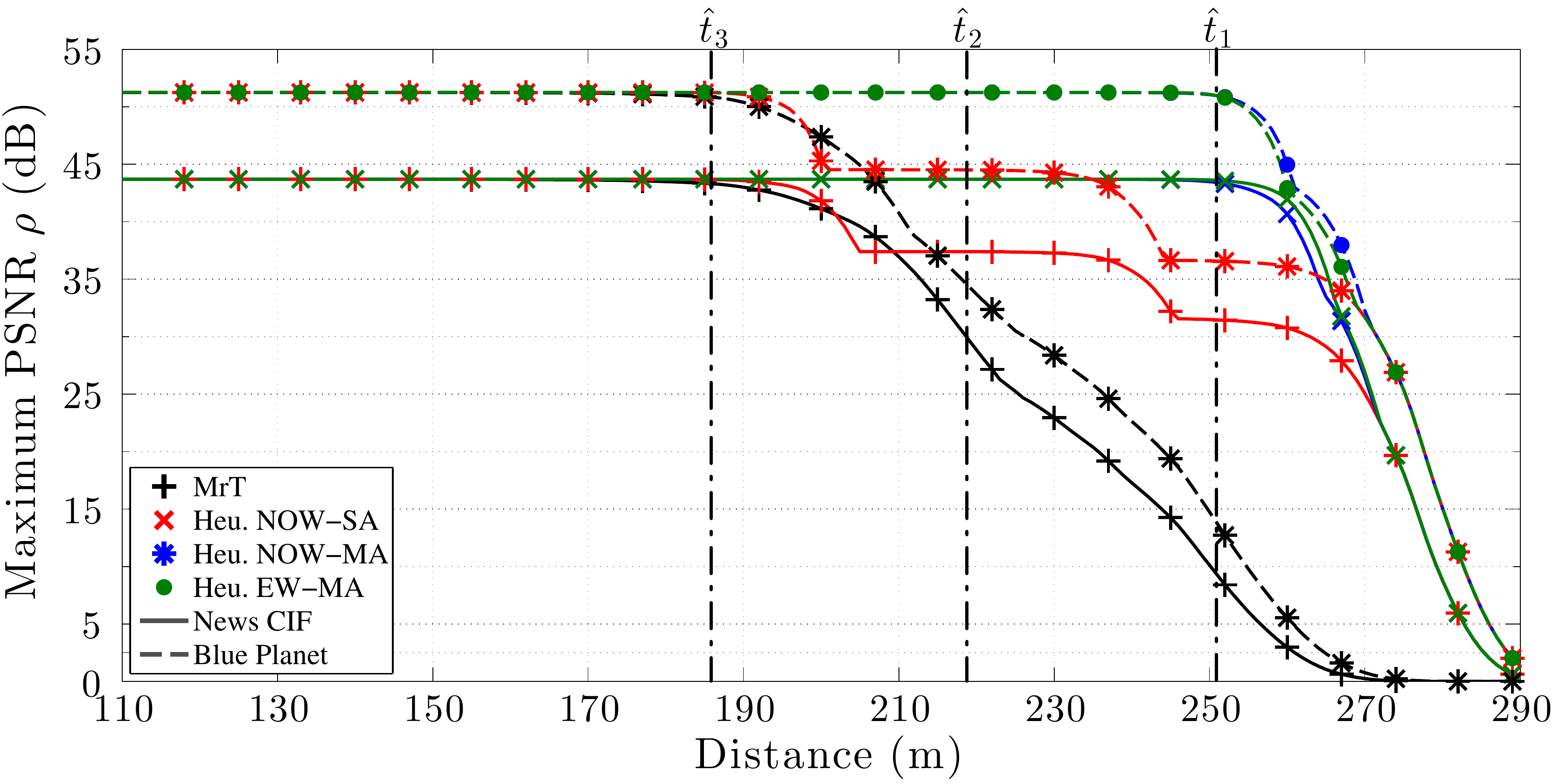}
\caption{Maximum PSNR associated with video streams News CIF and Blue Planet, for $q=2^8$.}
\label{fig.psnr_A_B}
\end{figure}

Fig.~\ref{fig.psnr_A_B} shows the value of $\rho$ associated with the streams News CIF and Blue Planet, as a function of distance from the centre of the cell, for $q = 2^8$. For the sake of comparison, we also report the performance of MrT even if it does not depend on the value of $q$. We recall from Fig.~\ref{fig.footprint} that the performance gap, in terms of the value of $\tau$, between the heuristic \mbox{NOW-SA/NOW-MA} and EW-MA solutions is small ($2$ TBs). As expected, the heuristic NOW-MA solution provides a service coverage that overlaps with that given by the heuristic EW-MA, in the case of both video streams. We can thus conclude that NOW-MA and EW-MA strategies perform similarly both in terms of resource footprint and service coverage, for large value of $q$. Furthermore, even though the NOW-SA approach is characterized by the same resource footprint of NOW-MA, the achieved service coverage still diverges from that of NOW-MA and EW-MA. Once more, this performance gap is caused by the fact that the NOW-SA approach cannot exploit user heterogeneity. Finally, we remark that, also in this case, all the proposed allocation models meet the required coverage constraints.

We demonstrated that the proposed resource allocation frameworks fulfil the desired goals set in Section~\ref{sec:intro}, namely (i) to ensure the desired QoS levels to at least a target fraction of users, and (ii) to minimize the required number of TB transmissions. In particular, we established that each proposed framework ensures a service coverage, which not only meets the target performance but also outperforms the service coverage provided by the considered MrT strategy. 
A fact that should be kept in mind is that the increased service coverage factor depends on the user propagation conditions and the bitrate of each service layer. However, if the transmitted multicast services have similar bitrates, the increased service coverage can be directly translated into an enlargement of the inter-site-distance or a better placement of the base stations.

\section{Conclusions and Future Research Directions}\label{sec:conclusions}
In this paper, we studied a general system model in which a source node uses point-to-multipoint (PtM) transmission to multicast a layered message to a group of users. The number of consecutive layers recovered by a user determines the QoS level of that user. In order to improve communication reliability, we considered Random Linear Network Coding (RLNC) and we investigated two different implementations, which are suitable for layered source messages: the Non-Overlapping Window (NOW-RLNC) and the Expanding Window (EW-RLNC) schemes. We derived accurate closed-form expressions for the probability of recovering a predetermined set of consecutive message layers for both NOW-RLNC and EW-RLNC and we used these expressions to assess the user QoS. To maintain the generality of the system model and facilitate its extendibility to 4G and next-generation standards, we assumed that a layered source message can be transmitted over multiple orthogonal communication subchannels.

Based on this assumption, we developed resource allocation frameworks which aim to minimize the overall number of coded packet transmissions. The proposed frameworks allocate coded packets of the same layer or the same expanding window either to a single or to multiple subchannels; we called the former pattern Separated Allocation (SA) while the latter pattern Mixed Allocation (MA). A key point in the formulation of the resource allocation problems is that the derived solutions ensure that predetermined fractions of users can achieve the desired QoS with at least a target probability. We explained that both SA and MA are computationally complex integer problems but we proposed heuristic strategies which are capable of obtaining good-quality solutions in a finite number of steps.

As a case study, we presented a possible integration of the RLNC-based schemes into the standard \mbox{LTE-A} Media Access Control (MAC) layer and the adaptation of the developed resource allocation frameworks to \mbox{LTE-A} systems. In addition, we described how the resulting modified MAC layer can be used to efficiently deliver a layered multimedia stream compliant with the H.264/SVC  standard over an \mbox{LTE-A} network that operates in the Single-Cell eMBMS mode.

In order to investigate the performance of the proposed schemes, we referred to an \mbox{LTE-A} network scenario defined by 3GPP to benchmark urban cellular network deployments. Furthermore, we considered two video traces -- one of low bitrate and the other of high bitrate -- both of which are publicly available for network performance evaluation. The first part of our investigation compared heuristic solutions to solutions obtained by directly solving the optimization problems and established that our proposed heuristic strategies indeed produce good-quality solutions. In the second part of our analysis, we demonstrated that both NOW and EW schemes can offer the same quality of service, in terms of PSNR, as conventional multi-rate transmission (MrT) but over a much longer distance. For a $99$\% probability of recovering the base video layer, we showed that the proposed strategies can achieve a coverage that is greater than that of a conventional MrT strategy by a factor of at least $1.35$. Furthermore, we unveiled that EW-MA can achieve similar coverage to that of \mbox{NOW-SA} and \mbox{NOW-MA} but at a notable resource advantage when binary network coding is used. More specifically, EW-MA can reduce packet transmissions by $28$\% and $17$\% for the case of the considered low and high bitrate streams, respectively. Nevertheless, we clarified that as the field size of network coding increases, the NOW and EW schemes perform similarly.

Future research directions involve the optimization of the sparsity of RLNC as well as the definition of different optimization objectives. In this paper, we employed the classic implementation of RLNC, where coding coefficients are randomly selected over a finite field. It is well known from the literature that the coding coefficient selection can be biased in order to increase the probability of selecting a zero coefficient. We understand that, as the sparsity of a coding vector increases, the RLNC decoding complexity decreases. However, the more zero coefficients a coding vector has, the higher the probability is that a user receives linearly dependent coded packets. Owing to the lack of a theoretical characterization of the tradeoff between sparsity and decoding complexity, we will strive to reinterpret both the NOW-RLNC and EW-RLNC approaches. The resulting theoretical characterization will allow us to jointly optimize transmission parameters and the sparsity of RLNC.

\bibliographystyle{IEEEtran}
\bibliography{IEEEabrv,biblio}




\end{document}